\documentclass[journal]{IEEEtran}

\usepackage{glossaries}
\makeglossaries
\usepackage{makecell} 
\usepackage{cite}
\usepackage{graphicx}
\usepackage{algorithm}
\usepackage{algorithmicx}
\usepackage{algpseudocode}
\usepackage{array}
\usepackage{breakurl}
\usepackage[table]{xcolor}
\usepackage{booktabs} 

\usepackage{balance}

\newcommand{\myFontStages}{\fontfamily{cmss}\selectfont}

\newcommand{\mpstat}{{\fontfamily{lmss}\selectfont mpstat}}

\newcommand{\perf}{{\fontfamily{lmss}\selectfont perf}}

\newcommand{\qdisc}{{\fontfamily{lmss}\selectfont qdisc}}

\newcommand{\ethtool}{{\fontfamily{lmss}\selectfont ethtool}}

\newcommand{\ftrace}{{\fontfamily{lmss}\selectfont ftrace}}

\newcommand{\bcmgenet}{{\fontfamily{lmss}\selectfont bcmgenet}}

\newcommand{\netdev}{{\fontfamily{lmss}\selectfont netdev}}

\newcommand{\iwlwifi}{{\fontfamily{lmss}\selectfont iwlwifi}}

\newcommand{\maceighteleven}{{\fontfamily{lmss}\selectfont mac80211 }}

\newcommand{\initXmit}{{\myFontStages Init Xmit}}
\newcommand{\initKsoftirqd}{{\myFontStages Init Ksoftirqd}}
\newcommand{\initRXIRQ}{{\myFontStages Init RX-\gls{IRQ}}}
\newcommand{\RXIRQ}{{\myFontStages RX SoftIRQ}}
\newcommand{\pollFunction}{{\myFontStages Poll Function}}
\newcommand{\ipStack}{{\myFontStages IP Stack}}
\newcommand{\initTXIRQ}{{\myFontStages Init TX-IRQ}}
\newcommand{\TXIRQ}{{\myFontStages TX SoftIRQ}}
\newcommand{\TXQdisc}{{\myFontStages TX Qdisc}}
\newcommand{\TXReclaim}{{\myFontStages TX Reclaim}}

\newacronym{SNAT}{SNAT}{Source NAT}
\newacronym{NAT}{NAT}{Network Address Translation}

\newacronym{TID}{TID}{Traffic Identifier}
\newacronym{MPDU}{MPDU}{MAC Protocol Data Units}
\newacronym{MSIX}{MSIX}{Message Signaled Interrupts-Extended}
\newacronym{ISR}{ISR}{Interrupt Service Routine}
 
\newacronym{L1-DCache}{L1-DCache}{Layer-1 Data Cache}

\newacronym{FIB}{FIB}{Forwarding Information Base}

\newacronym{SKB}{SKB}{socket buffer}

\newacronym{FIFO}{FIFO}{First-In First-Out}

\newacronym{SBC}{SBC}{Single Board Computer}
\newacronym{MTU}{MTU}{Maximum Transmission Unit}

\newacronym{PU}{PU}{Processing Unit}
\newacronym{ML}{ML}{Machine Learning}

\newacronym{TTL}{TTL}{Time to Live}

\newacronym{IRQ}{IRQ}{Interrupt Request}
\newacronym{TH}{TH}{Top Half}
\newacronym{BH}{BH}{Bottom Half}

\newacronym{PCI}{PCI}{Peripheral Component Interconnect}

\newacronym{NAPI}{NAPI}{New API}

\newacronym{RIC}{RIC}{Receive Interrupt Coalescing}
 
\newacronym{CM4}{CM4}{Compute Module 4}

\newacronym{MMIO}{MMIO}{Memory-Mapped I/O}

\newacronym{IoT}{IoT}{Internet of Things}

\newacronym{SCU}{SCU}{Santa Clara University}
\newacronym{UTM}{UTM}{University Technology Malaysia}
\newacronym{SES}{SES}{Summer Engineering Seminar}

\newacronym{RTOS}{RTOS}{Real-Time Operating System}

\newacronym{INT}{INT}{In-band Network Telemetry}

\newacronym{QS}{QS}{Queue Size}

\newacronym{QSA}{QSA}{Queue Size All}

\newacronym{DMA}{DMA}{Direct Memory Access}

\newacronym{MCS}{MCS}{Modulation and Coding Scheme}

\newacronym{E2W}{E2W}{Ethernet to WiFi}
\newacronym{W2E}{W2E}{WiFi to Ethernet}

\newglossaryentry{RTT}
{
  name={RTT},
  description={Round-Trip Time},
  first={\glsentrydesc{RTT} (\glsentrytext{RTT})},
  plural={RTTs},
  firstplural={\glsentrydesc{RTT}s (\glsentryplural{RTT})}
}

\newacronym{MQTT}{MQTT}{Message Queue Telemetry Transport}

\newacronym{API}{API}{Application Programming Interface}

\newacronym{WSN}{WSN}{Wireless Sensor Networks}

\newacronym{EDF}{EDF}{Earliest Deadline First}

\newacronym{MAE}{MAE}{Mean Absolute Error}

\newacronym{MLP}{MLP}{Multilayer perceptrons}
\newacronym{RFR}{RFR}{Random Forest Regressor}
\newacronym{GBR}{GBR}{Gradient Boosting Regressor}
\newacronym{ABR}{ABR}{Adaptive Boosting Regressor}
\newacronym{ETR}{ETR}{Extra Trees Regressor}
\newacronym{LSTM}{LSTM}{Long Short-Term Memory}
\newacronym{HBR}{HBR}{Histogram-Based Gradient Boosting Regressor}
\newacronym{NN}{NN}{Neural Networks}
\newacronym{RNN}{RNN}{Recurrent Neural Networks}

\newacronym{HTB}{HTB}{Hierarchical Token Bucket}

\newacronym{H2H}{H2H}{Humand to Human}
\newacronym{M2H}{M2H}{Machine to Human}
\newacronym{M2M}{M2M}{Machine to Machine}

\newacronym{MAC}{MAC}{Medium Access Control}
\newacronym{CW}{CW}{Contention Window}

\newacronym{WUR}{WUR}{Wake-up Radio}

\newacronym{RAT}{RAT}{Radio Access Technology}

\newacronym{CAGR}{CAGR}{Compound Annual Growth Rate}

\newacronym{RSSI}{RSSI}{Received Signal Strength Indicator}
\newacronym{SNR}{SNR}{Singal-to-Noise Ratio}

\newacronym{BLE}{BLE}{Bluetooth Low Energy}

\newacronym{NFC}{NFC}{Near Field Communication}

\newacronym{WEP}{WEP}{Wired Equivalent Privacy }
\newacronym{WPA}{WPA}{Wireless Protected Access}
\newacronym{PMK}{PMK}{Pairwise Master Key}
\newacronym{PTK}{PTK}{Pairwise Transition Keys}
\newacronym{GTK}{GTK}{Group Temporal Key}
\newacronym{EAP}{EAP}{Extensible Authentication Protocol}
\newacronym{RRM}{RRM}{Radio Resource Management}
\newacronym{BSSTM}{BSSTM}{BSS Transition Management}
\newacronym{BSS}{BSS}{Basic Service Set}
\newacronym{SSID}{SSID}{Service Set Identifier}
\newacronym{BSSID}{BSSID}{BSS Identifier}
\newacronym{CSA}{CSA}{Channel Switching Announcement}
\newacronym{DCF}{DCF}{Distributed Coordination Function}
\newacronym{EDCF}{EDCF}{Enhanced Distributed Coordination Function}
\newacronym{AIFS}{AIFS}{Arbitration Inter-Frame Space}

\newacronym{QoS}{QoS}{Quality of Service}
\newacronym{ToS}{ToS}{Type of Service}

\newacronym{TSF}{TSF}{Time Synchronization Function}
\newacronym{AC}{AC}{Access Category}

\newacronym{SMP}{SMP}{Symmetric Multiprocessing}

\newacronym{NTP}{NTP}{Time Synchronization Protocol}

\newacronym{PSM}{PSM}{Power Save Mode}
\newacronym{PSP}{PSP}{Power Save Poll}
\newacronym{PSP-N}{PSP-N}{Power Save Poll with Null}
\newacronym{CAM}{CAM}{Continuously Active Mode}
\newacronym{AID}{AID}{Association ID}
\newacronym{APSM}{APSM}{Adaptive PSM}
\newacronym{APSD}{APSD}{Automatic Power Save Delivery}
\newacronym{U-APSD}{U-APSD}{Unscheduled-Automatic Power Save Delivery}
\newacronym{S-APSD}{S-APSD}{Scheduled-Automatic Power Save Delivery}
\newacronym{EOSP}{EOSP}{End of Service Period}
\newacronym{SP}{SP}{Service Period}
\newacronym{TXOP}{TXOP}{Transmit Opportunity}
\newacronym{BI}{BI}{Beacon Interval}

\newacronym{DBS}{DBS}{Disassociation-Based Steering}

\newacronym{SNMP}{SNMP}{Simple Network Management Protocol}

\newacronym{ODL}{ODL}{OpenDayLight} 
\newacronym{OVS}{OVS}{Open vSwitch}
\newacronym{OVSDB}{OVSDB}{Open vSwitch Database}
\newacronym{SDN}{SDN}{Software Defined Networking}

\newacronym{VM}{VM}{Virtual Machine}

\newacronym{NF}{NF}{Network Function}
\newacronym{VNFM}{VNFM}{VNF Manager}
\newacronym{VNF}{VNF}{Virtual Network Function}
\newacronym{NNF}{NNF}{Native Network Function}
\newacronym{NFVO}{NFVO}{NFV Orchestrator}
\newacronym{VIM}{VIM}{Virtualized Infrastructure Manager}

\newacronym{PCP}{PCP}{Priority Code Point}

\newacronym{NUMA}{NUMA}{Non-Uniform Memory Access}

\newacronym{CPE}{CPE}{Customer Premise Equipment}

\newacronym{DSCP}{DSCP}{Differentiated Services Code Point}

\newacronym{vSG}{vSG}{Virtual Subscriber Gateway}

\newacronym{RPS}{RPS}{Receive Packet Steering}
\newacronym{XPS}{XPS}{Transmission Packet Steering}

\newacronym{IPS}{IPS}{Intrusion Prevention System}
\newacronym{IDS}{IDS}{Intrusion Detection System}
\newacronym{VPN}{VPN}{Virtual Private Network}

\newacronym{RGW}{RGW}{Residential Gateway}
\newacronym{vRGW}{vRGW}{Virtual RGW}

\newacronym{VF}{VF}{Virtual Function}
\newacronym{PF}{PF}{Physical Function}

\newacronym{AEA}{AEA}{AP's eBPF Agent}
\newacronym{SEA}{SEA}{STA's eBPF Agent}

\newacronym{RPi}{RPi}{Raspberry Pi}

\newacronym{ARP}{ARP}{Address Resolution Protocol}

\newacronym{TIM}{TIM}{Traffic Indication Map}
\newacronym{DTIM}{DTIM}{Delivery Traffic Indication Message}

\newglossaryentry{A-MPDU}
{
  name={A-MPDU},
  description={Aggregated MPDU},
  first={\glsentrydesc{A-MPDU} (\glsentrytext{A-MPDU})},
  plural={A-MPDUs},
  firstplural={\glsentrydesc{A-MPDU}s (\glsentryplural{A-MPDU})}
}

\newacronym{HPC}{HPC}{High Performance Computing}

\newacronym{Mbps}{Mbps}{Mega Per Second}
\newacronym{Gbps}{Gbps}{Gigabits Per Second}
\newacronym{Tbps}{Tbps}{Terabits Per Second}

\newacronym{AP}{AP}{Access Point}

\newacronym{TWT}{TWT}{Target Wake Time}
\newacronym{OPS}{OPS}{Opportunistic Power Save}
\newacronym{PCF}{PCF}{Point Coordination Function}
\newacronym{PIFS}{PIFS}{PCP Inter Frame Space}
\newacronym{DIFS}{DIFS}{Distributed Inter Frame Space}
\newacronym{BSR}{BSR}{Buffer Status Report}
\newacronym{OCW}{OCW}{OFDM Contention Window}
\newacronym{OFDMA}{OFDMA}{Orthogonal Frequency-Division Multiple Access}
\newacronym{OBO}{OBO}{OFDMA Back-off}
\newacronym{RU}{RU}{Resource Unit}

\newacronym{QTP}{QTP}{Quite Time Period}
\newacronym{AU}{AU}{Airtime Utilization}

\newacronym{LVAP}{LVAP}{Light Virtual AP}
\newacronym{EDCA}{EDCA}{Enhanced Distributed Channel Access}

\newacronym{HCCA}{HCCA}{Hybrid Controlled Channel Access}

\newacronym{WOL}{WOL}{Wake-on-Lan}

\newacronym{DC}{DC}{Datacenter}
\newacronym{CO}{CO}{Central Office}

\newacronym{CORD}{CORD}{Central Office Re-architected as a Datacenter}
\newacronym{MEC}{MEC}{Mobile Edge Computing}
\newacronym{ASP}{ASP}{Application Service Provider}
\newacronym{ES}{ES}{Edge Service}
\newacronym{NSP}{NSP}{Network Service Provider}

\newacronym{NV}{NV}{Network Virtualization}
\newacronym{NFV}{NFV}{Network Function Virtualization}
\newacronym{NFVI}{NFV}{NFV Infrastructure}
\newacronym{MANO}{MANO}{Management and Orchestration}

\newacronym{RX}{RX}{receive}
\newacronym{TX}{TX}{transmit}

\newacronym{NIC}{NIC}{Network Interface Card}
\newacronym{vNIC}{vNIC}{virtual NIC}
\newacronym{pNIC}{pNIC}{physical NIC}

\newacronym{TCAM}{TCAM}{Tenary Content Addressable Memory}
\newacronym{RSS}{RSS}{Receive Side Scaling}

\newacronym{BQL}{BQL}{Byte Queue Limits}

\newacronym{GRO}{GRO}{Generic Receive Offload}
\newacronym{GSO}{GSO}{Generic Segmentation Offload}

\newacronym{BDP}{BDP}{Bandwidth Delay Product}

\newacronym{DPDK}{DPDK}{Data Plane Development Kit}
\newacronym{VPP}{VPP}{Vector Packet Processing}

\newacronym{UIO}{UIO}{User-space I/O}

\newacronym{EMC}{EMC}{Exact Match Cache}

\newacronym{SR-IOV}{SR-IOV}{Single Root-Input/Output Scaling}
\newacronym{PMD}{PMD}{Poll Mode Driver}

\newacronym{TSS}{TSS}{Tuple Space Search}
\newacronym{dpcls}{dpcls}{data path classifier}

\newacronym{FDK}{FDK}{Fog Development Kit}
\newacronym{RAA}{RAA}{Resource Allocation Algorithm}

\newacronym{TSN}{TSN}{Time Sensitive Networking}
\newacronym{AFDX}{AFDX}{avionics full-duplex switched Ethernet}
\newacronym{CAN}{CAN}{Controller Area Network}

\newacronym{OCI}{OCI}{Open Carrier Interface }

\newacronym{CBR}{CBR}{constant bit rate}

\newacronym{EAPS-E}{EAPS-E}{EAPS with Early wake-up}
\newacronym{EAPS-L}{EAPS-L}{EAPS with Late wake-up}
\newacronym{EAPS-M}{EAPS-M}{EAPS with Mid wake-up}

\newacronym{ECDF}{ECDF}{Empirical Cumulative Distribution Function}

\newacronym{STA}{STA}{station}
\newacronym{BSRP}{BSRP}{Buffer State Report Poll}
\newacronym{MSDU}{MSDU}{MAC Service Data Unit}
\newacronym{COTS}{COTS}{Commercial Off-The-Shelf}

\newacronym{EWU}{EWU}{Early Wake-UP}
\newacronym{LWU}{LWU}{Late Wake-UP}
\newacronym{MWU}{MWU}{Mid Wake-UP}

\newacronym{DP}{DP}{Data Path}
\newacronym{FCC}{FCC}{Federal Communications Commission}
\newacronym{MPTCP}{MPTCP}{MultiPath TCP}

\newacronym{HAL}{HAL}{Hardware Abstraction Layer}

\newacronym{NOTP}{FLIP}{Name Of The Paper}
\newacronym{EDC}{EDC}{Event-based Data Collection}
\newacronym{PDC}{PDC}{Polling-based Data Collection}
\newacronym{EPDC}{EPDC}{Event and Polling-based Data Collection}

\newacronym{IMI}{IMI}{Inter-Measurement Interval}
\newacronym{RBL}{RBL}{regular background load}
\newacronym{HBL}{HBL}{high background load}
\newacronym{NOTP-NL}{NOTP-NL}{\gls{NOTP} with Netlink}
\newacronym{NOTP-MNL}{NOTP-MNL}{\gls{NOTP} with memory-mapped netlink}
\newacronym{CPUDED-D}{CPUDED-D}{Dedicated CPU disabled}
\newacronym{CPUDED-E}{CPUDED-E}{Dedicated CPU enabled}

\newacronym{ACVO}{${AC}_{VO}$}{AC VOice}
\newacronym{ACVI}{${AC}_{VI}$}{AC VIdeo}
\newacronym{ACBE}{${AC}_{BE}$}{AC Best Effort}
\newacronym{ACBK}{${AC}_{BK}$}{AC BacKground}

\newacronym{BT}{BT}{Background Traffic}

\newacronym{XDP}{XDP}{eXpress Data Path}
\newacronym{BMv2}{BMv2}{Behavioral Model}

\newacronym{PEM}{PEM}{Passive Energy-state Monitor}
\newacronym{ESM}{ESM}{Energy State Monitor}
\newacronym{NSM}{NSM}{Network State Monitor}

\newacronym{BTG}{BTG}{Background Traffic Generator}
\newacronym{EAPOL}{EAPOL}{Extensible Authentication Protocol Over LAN}
\newacronym{LAN}{LAN}{Local Area Network}
\newacronym{MAN}{MAN}{Metropolitan Area Network}
\newacronym{TBTT}{TBTT}{Target Beacon Transmission Time}

\newacronym{NTEC}{NTEC}{Network Throughput with respect to Energy Consumption}
\newacronym{eBPF}{eBPF}{extended Berkeley Packet Filter}
\newacronym{CD}{CD}{Cumulative Delay}
\newacronym{ATU}{ATU}{Awake-Time Utilization}

\newacronym{AWS}{AWS}{Amazon Web Services}
\newacronym{DDoS}{DDoS}{Distributed Denial-of-Service}
\newacronym{CU}{CU}{Channel Utilization}
\newacronym{DL}{DL}{downlink}
\newacronym{UL}{UL}{uplink}

\newacronym{NSM-U}{NSM-U}{NSM user-space}
\newacronym{NSM-K}{NSM-K}{NSM kernel-space}

\newacronym{ADC}{ADC}{Analog-to-Digital Converter}
\newacronym{SPI}{SPI}{Serial Peripheral Interface}

\newacronym{CCA}{CCA}{Clear Channel Assessment}

\makeatletter
\def\verbatim@font{\normalfont\ttfamily\fontfamily{lmtt}\selectfont}
\makeatother

\ifCLASSINFOpdf
\else
\fi

\def\BibTeX{{\rm B\kern-.05em{\sc i\kern-.025em b}\kern-.08em
    T\kern-.1667em\lower.7ex\hbox{E}\kern-.125emX}}

\newcounter{observation}
\newenvironment{observation}[1][]{%
  \refstepcounter{observation}
  \vspace{7pt}
  \noindent
  \textbf{Observation \theobservation:} \textit{#1}%
  \itshape
}{%
  \normalfont
  \vspace{5pt}
}

\usepackage[inline,shortlabels]{enumitem}

\begin{document}

\title{Understanding and Enhancing Linux Kernel-based Packet Switching on WiFi Access Points}

\author{\IEEEauthorblockN{Shiqi Zhang, Mridul Gupta, and Behnam Dezfouli}
\\
\IEEEauthorblockA{\small Internet of Things Research Lab, Department of Computer Science and Engineering, Santa Clara University, USA}
\\
\texttt{\small\{szhang9, magupta, bdezfouli\}@scu.edu\quad}
}

\maketitle

\begin{abstract}
As the number of WiFi devices and their traffic demands continue to rise, the need for a scalable and high-performance wireless infrastructure becomes increasingly essential. Central to this infrastructure are WiFi Access Points (APs), which facilitate packet switching between Ethernet and WiFi interfaces. Despite APs' reliance on the Linux kernel's data plane for packet switching, the detailed operations and complexities of switching packets between Ethernet and WiFi interfaces have not been investigated in existing works. This paper makes the following contributions towards filling this research gap. Through macro and micro-analysis of empirical experiments, our study reveals insights in two distinct categories. Firstly, while the kernel's statistics offer valuable insights into system operations, we identify and discuss potential pitfalls that can severely affect system analysis. For instance, we reveal the implications of device drivers on the meaning and accuracy of the statistics related to packet-switching tasks and processor utilization. Secondly, we analyze the impact of the packet switching path and core configuration on performance and power consumption. Specifically, we identify the differences in Ethernet-to-WiFi and WiFi-to-Ethernet data paths regarding processing components, multi-core utilization, and energy efficiency. We show that the WiFi-to-Ethernet data path leverages better multi-core processing and exhibits lower power consumption.

\end{abstract}


\IEEEpeerreviewmaketitle

\glsresetall

\section{Introduction}

The advent of new applications and services, particularly those involving high-definition streaming, edge and cloud computing, and increasingly sophisticated \gls{IoT} devices, necessitates the enhancement of WiFi technology in terms of rate, reliability, and scalability \cite{reshef2022future}. 
Parallel to this need for speed and efficiency, there is a significant increase in the number of WiFi \glspl{AP} (henceforth referred to as \glspl{AP}), which are mainly responsible for implementing a data plane for switching packets between their Ethernet and WiFi \glspl{NIC}. 
Statistics show that the gigabit  \gls{AP} market alone is projected to increase at a compounded annual growth rate of about 32.3\% from 2024 to 2034 \cite{wifi_ap_market_2024}. This upsurge is a response to the expanding reach of internet connectivity and the need for high-speed and reliable coverage across various spaces.

Compared to enterprise and datacenter switches, which typically implement their dataplane functions in hardware for high-speed processing, \glspl{AP} often rely on the Linux kernel for packet switching tasks \cite{hoiland2017ending,sheth2021monfi,eaps_jay}.
This architectural difference is primarily due to the differing packet switching rates: modern enterprise switches handle immense data traffic, often exceeding several Tbps, whereas \glspl{AP} generally manage lower rates, typically in the range of Mbps or Gbps. This lower switching rate is due to their focus on providing wireless network access rather than core data routing. Consequently, using the Linux kernel for software-based packet switching in \glspl{AP} offers a cost-effective and flexible solution that meets their data throughput requirements.

Given the rising complexity of operations and the increasing number of \glspl{AP} deployed in various settings, understanding and enhancing the performance of packet switching on these devices is becoming crucial.
The existing research on software packet switching, however, primarily focuses on high-performance environments including resourceful servers \cite{emmerich2014performance,powell2020fog,cai2021understanding,okelmann2021adaptive}. 
Additionally, given the superior performance of kernel-bypass methods like \gls{DPDK}, existing studies  emphasize these technologies \cite{zhang2019comparing,gallardo2016performance,okelmann2021adaptive,chen2021predictable} for packet switching between Ethernet \glspl{NIC} and user-space components such as \glspl{VM} and containers.
In contrast, \glspl{AP} have distinct requirements, primarily performing packet switching between Ethernet and WiFi \glspl{NIC}.
Moreover, to understand and analyze the data plane of \glspl{AP}, it is essential to establish a systematic approach by leveraging high-level statistics provided by the kernel and the detailed insights from specific components managing tasks along the data path from one \gls{NIC} to another.

In this paper, we study and analyze the operation and performance of two data paths, namely \gls{E2W} and \gls{W2E} packet switching paths, on WiFi \glspl{AP}, focusing on two main questions: 
{(i)} \textit{How and to what extent can the statistics provided by Linux and its performance monitoring tools be used to understand packet switching operation and performance?}
{(ii)} \textit{What are the differences and their causes in processing resources and power consumption between \gls{E2W} and \gls{W2E} packet switching?}
To address these questions, we first provide an overview of the primary operations and stages of packet switching in the Linux kernel. Then, we empirically study and analyze packet switching operations on an ARM-based platform, which utilizes a processor similar to those used in commercial \glspl{AP}. 
Integral to this empirical analysis, we perform both macro and micro-analysis of packet switching operations to understand the performance characteristics and resource consumption patterns unique to WiFi \glspl{AP}.

{Our main findings are as follows.}
\begin{enumerate*}[label=(\roman*)]
\item \textbf{While Linux provides statistics about the number of \glspl{IRQ} and SoftIRQs, the meanings of these parameters are highly affected by various system configuration settings.} 
For instance, we show that the number of RX SoftIRQs is influenced not only by the processing of incoming packets but also by the number of TX \glspl{IRQ}.
Also, while Linux's networking subsystem specifies a protocol for switching between \gls{NAPI} (polling) and \gls{IRQ} operations, some drivers override this protocol, therefore changing the meaning of the statistics provided.
\item \textbf{The processor cycles consumed by SoftIRQ processing may not be accurately accounted for and reflected in processor utilization statistics.}
Specifically, the SoftIRQ instances handled closely following the \gls{TH} interrupt processing and before running a \verb|ksoftirqd| thread may be considered part of the \verb|idle| processor utilization instead of SoftIRQ utilization.
We observe this behavior for the Ethernet interface of the platform under test.
\item \textbf{Even when the two \glspl{NIC} are assigned to separate cores, packet switching in the \gls{E2W} path does not utilize multi-core processing.} This limitation can result in a throughput drop if the core assigned to the Ethernet \gls{NIC} cannot handle the maximum supported throughput of the two \glspl{NIC}. In contrast, packet switching in the \gls{W2E} path inherently utilizes two cores.
\item \textbf{The processing load and power consumption of the \gls{E2W} path are higher than those of the \gls{W2E} path. However, as the throughput reaches the maximum supported levels, the differences in efficiency between the two paths reduce.}
\end{enumerate*}

The rest of this paper is organized as follows. Section \ref{linux_pkt_switch_overview} provides an overview of packet processing stages in the Linux kernel. Testbed components and the collection of performance metrics are detailed in Section \ref{section_testbed}. We present a macro-analysis of packet switching metrics in Section \ref{macro_E2W}. In Section \ref{micro_comp_directions}, we provide an in-depth, micro-analysis of the packet switching operations. Section \ref{related_work} presents related work and future directions. We conclude the paper in Section \ref{sec_conclusion}.

\glsresetall

\section{An Overview of Linux Kernel Packet Switching}
\label{linux_pkt_switch_overview}

\begin{figure*}[!ht]
    \centering
    \includegraphics[width=0.95\linewidth]{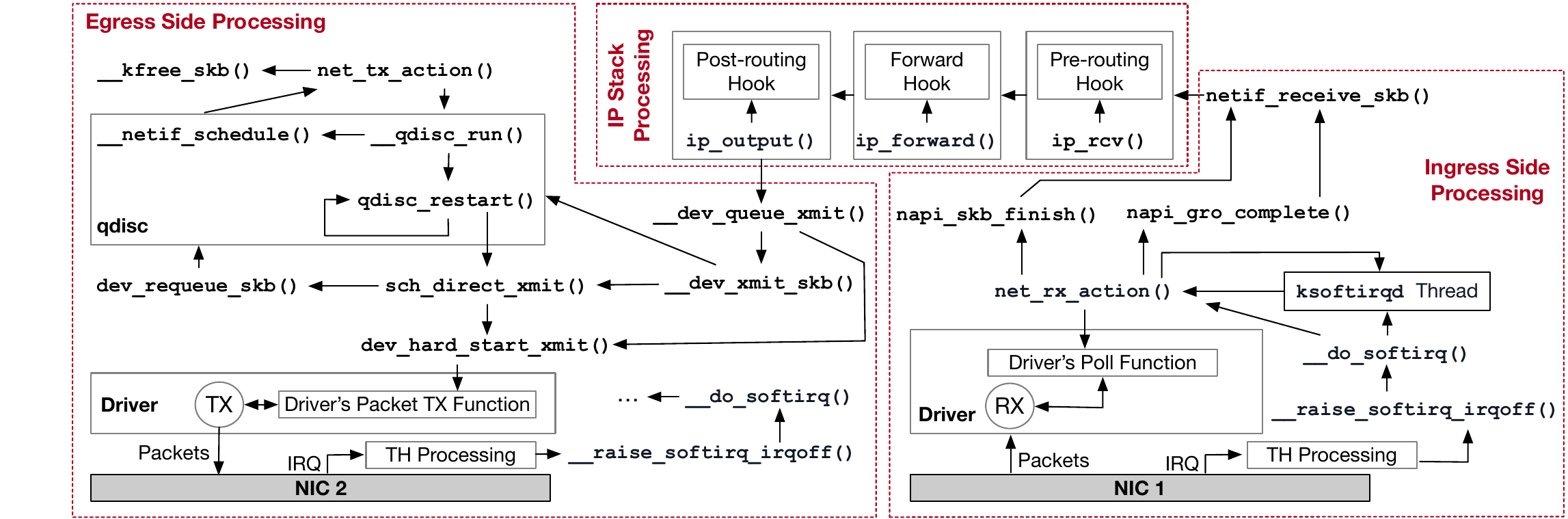}
    \caption{The main steps of packet switching. 
    Here, we assume the ingress interface is  \gls{NIC} 1 and the egress interface is \gls{NIC} 2.}
    \label{fig:pkt_switching_steps}
\end{figure*}

In this section, we provide an overview of the steps involved in the Linux kernel's data path for switching packets received on an ingress \gls{NIC} to an egress \gls{NIC}, based on our analysis of the latest Linux kernel source code\footnote{At the time of writing this paper, the latest available kernel version is 6.7.}. 
Figure \ref{fig:pkt_switching_steps} illustrates the major steps in the process of packet switching from NIC 1 (ingress) to NIC 2 (egress).

\subsection{Ingress Side Processing}
\label{sec:reviews_rx_side}
\textbf{NIC to RX buffer.}
When a network packet arrives at a \gls{NIC}, it is first stored in the \gls{NIC}'s internal buffers. 
The \gls{NIC} then uses \gls{DMA} to transfer packets to the system's main memory, specifically into the buffers managed by the driver, updating the \gls{RX} buffer's pointers in the process.
The \gls{RX} buffer is a ring queue, facilitating producer (\gls{NIC}) and consumer (driver) operation.
This action triggers an \gls{IRQ}, invoking the corresponding interrupt handler, often referred to as the \glsfirst{TH}.
The \gls{TH} acknowledges the interrupt, performs preliminary handling, and then schedules further processing in a deferred manner, referred to as the \gls{BH}.

Generating an interrupt for each packet reception results in a high interrupt processing overhead.
\gls{NAPI} addresses this problem by utilizing a \textit{polling} mode, where the driver periodically checks if new descriptors have been used for packet reception.
Within this method, processing a \gls{TH} involves calling the \verb|napi_schedule| function of the kernel to start the \gls{NAPI} for the RX buffer.
To this end, when processing the \gls{TH} for an \gls{IRQ}, the \verb|napi_schedule| function is called with a \verb|napi_struct| parameter that includes a pointer to the poll function of the driver.
Afterwards, the \gls{TH} calls the \verb|__raise_softirq_irqoff| function to raise a SoftIRQ, specifically an RX SoftIRQ designated for receiving packet processing.
At this point, \gls{BH} execution of ingress processing starts.

\textbf{SoftIRQ execution.} 
If there is any pending SoftIRQ, the kernel's \verb|__do_softirq| function handles RX SoftIRQs by calling the \verb|net_rx_action| function, which processes the network packets.
The \verb|net_rx_action| function utilizes the per-core \verb|softnet_data| structure to access and iterate over the \verb|napi_struct| list. 
Each entry of this list represents a network device with packets ready to be processed. 
The function dequeues each structure in turn and runs their driver's poll functions. 
It is important to note that \textit{the SoftIRQ is run on the same core that ran the \gls{TH}, i.e., the core that received the \gls{IRQ}}.
It is also important to note that after processing the \gls{TH}, the \verb|__do_softirq| function can immediately run within the \gls{IRQ} context, subject to certain limits on the number and duration of the SoftIRQs called.
Once the limits are exceeded, if more SoftIRQs must be run, the \verb|__do_softirq| function can execute in the context of a specific kernel thread called \verb|ksoftirqd|.
We will elaborate on these limitations in Observation \ref{obsr:tx_irq}.


\textbf{Switching between \gls{NAPI} and IRQ.}
The contract between the \gls{NAPI} subsystem and device drivers includes an important aspect related to the deactivation of \gls{NAPI}.
Every time a driver's poll function is called, the number of processed packets is returned in the \verb|work_done| variable. 
If a driver's poll function consumes its full \verb|weight| allotment (set as 64 by default), the \gls{NAPI} state stays unchanged, and control is returned to the \verb|net_rx_action| loop, which may call the poll function again if the processing time limit allows.
Conversely, if the poll function does not use its entire weight, it must disable \gls{NAPI}. Subsequently, \gls{NAPI} will be reactivated upon receipt of the next IRQ; at this point, the driver's \gls{IRQ} handler is expected to invoke the \verb|napi_schedule| function.

\textbf{Delivery to IP stack.}
When a packet is not suitable for \gls{GRO} aggregation, it is processed individually. 
This occurs within the \verb|napi_skb_finish| function, which calls \verb|netif_receive_skb| to handle such packets.
When the packet must be processed with \gls{GRO}, and it is time to terminate the ongoing aggregation, the function \verb|napi_gro_complete| is invoked.
This function then calls \verb|netif_receive_skb| to process the aggregated packets individually.
In both cases, \verb|netif_receive_skb| is responsible for further packet processing and delivery to the IP stack.
In Section \ref{section_testbed}, we will highlight that in this paper we do not utilize \gls{GRO}.

\subsection{IP Stack Processing}
For IP packets, the function \verb|ip_rcv| is called.
This function calls the \textit{pre-routing} hook of the Netfilter subsystem.
If this hook does not drop the packet, the routing subsystem is consulted to determine its destination.
If the packet is meant for another system, the \verb|ip_forward| function is called to determine the packet's next hop by consulting the routing table.
Subsequently, the packet may be modified and evaluated against firewall rules within the \textit{forward} Netfilter hook. 
The kernel performs further routing decisions after passing the forward hook. This involves consulting the routing table to determine the next-hop address and the appropriate outgoing network interface for the packet. 
Afterward, various IP-level checks and updates occur. 
If the packet is modified (e.g., TTL is decremented), the IP header checksum is recalculated.
As the packet is almost ready to be sent out, it approaches the \textit{post-routing} hook. 
This hook is the last chance to inspect and modify the packet before it leaves the system.

\subsection{Egress Side Processing}
\label{sec:reviews_tx_side}

One of the main components of egress side processing is \qdisc{}.
In Linux, \qdisc{} (short for Queueing Discipline) is a mechanism used to control how packets are queued and transmitted.
The default \qdisc{} in most Linux distributions is pfifo\_fast, which implements a simple \gls{FIFO} queue with three bands of traffic priorities.
Another algorithm is FQ-CoDel, which is a modern \qdisc{} designed to combat bufferbloat by managing queue lengths and ensuring fair bandwidth distribution.
Although Ethernet \glspl{NIC} utilize a \qdisc{} on their egress direction, in Section \ref{macro_analysis} we will discuss the lack of \qdisc{} on the egress direction of WiFi \gls{NIC} and its implications on performance.

\textbf{IP stack to driver.} With this background about \qdisc{}, we now explain the operations of egress side processing.
After IP stack processing, the packet is passed to the \netdev{} (short for network device) subsystem using the \verb|__dev_queue_xmit| function.
If there is a \qdisc{} associated with the interface, this function passes the packet to the \verb|__dev_xmit_skb|.
Otherwise, the packet is directly added to the \gls{TX}  buffer (a ring queue) of the driver by using the \verb|dev_hard_start_xmit| function.

If there is a \qdisc{} associated with the interface, the function  \verb|__dev_xmit_skb| first acquires a lock on the \qdisc{}, and then checks if the packet can {bypass} the \qdisc{} under certain conditions, such as when the \qdisc{} is empty.
If these conditions are met, the function attempts to transmit the packet directly, via the \verb|sch_direct_xmit|, \textit{bypassing the usual queuing mechanisms for efficiency.}
If \verb|__dev_xmit_skb| cannot bypass the \qdisc{}, the packet is enqueued in the \qdisc{}.
Then, if the \qdisc{} is not running, the function \verb|__qdisc_run| is called to run it.
As long as there are packets in the \qdisc{} and the quota of running \qdisc{} has not been fully used, \verb|__qdisc_run| calls \verb|qdisc_restart| in a loop to dequeue packets from the \qdisc{} and send them to the \gls{TX} buffer.
When the quota is exhausted, the \verb|__netif_schedule| function is called to schedule a TX SoftIRQ, which is used for processing outgoing network packets by running the \verb|net_tx_action| function.
This function accesses the \verb|softnet_data| structure of \textit{the core to which the \gls{IRQ} of the egress \gls{NIC} is assigned affinity}, and checks if the TX buffer has any SKBs that must be freed (using the \verb|__kfree_skb| function).
Then, if there are more packets in the \qdisc{}, the function \verb|__netif_schedule| runs the \qdisc{} again.

To pass a packet to the driver, a lock is 
acquired by the \verb|dev_hard_start_xmit| function on the driver's \gls{TX} buffer to prevent concurrent access to the transmit queue by other cores, ensuring that the packet transmission process is thread-safe.
After attempting to transmit the packet, the transmission lock on the \gls{TX} buffer is released.
The function then checks if the transmission was complete. 
If the transmission was not successfully completed (e.g., if the network driver did not successfully validate SKB), the packet is requeued using \verb|dev_requeue_skb| to attempt transmission again later.

\textbf{Driver to \gls{NIC}.}
The driver allocates buffers in the system's RAM, from which the \gls{DMA} engine can read the data and transfer to the \gls{NIC}.
Once the \gls{DMA} is set up, the driver triggers the network device to initiate the transmission. 
This usually involves writing to specific device registers, indicating the readiness of a packet for transmission and providing the \gls{DMA}-prepared memory buffers' locations. 
Once the \gls{NIC} completes the packet transmission, it signals this completion by generating a TX \gls{IRQ} to inform the driver that the transmission has finished. 
The \gls{IRQ} results in calling \gls{TH} and then \gls{BH} to perform several critical post-transmission tasks. 
It starts by unmapping any \gls{DMA} mappings that were previously established, ensuring proper memory management and preventing resource leaks. 
Following this, the driver frees the memory buffers allocated for the packet, typically involving the deallocation of the associated SKBs.

\section{Testbed Components and Collection of Performance Metrics}
\label{section_testbed}

We use the \gls{RPi} version 4 based on the \gls{CM4}, which includes an ARM Cortex-A72 processor. This processor has four cores, denoted as Core 0 through Core 3. We chose this platform because of the similarity of its processor to those used in \gls{COTS} \glspl{AP} such as those described in \cite{Qualcomm_IPQ8074} and \cite{Qualcomm_QCS5430}.
The \gls{CM4} module uses the BCM54210PE \cite{BCM54210} Ethernet controller, which utilizes the Broadcom Genet driver implemented as the \bcmgenet{} kernel module. Additionally, the \gls{CM4} module integrates a PCIe 2.0 x1 host controller, which we employed to connect an Intel AX210 WiFi 6E (802.11ax) \gls{NIC} to the platform. 
We enabled WPA3 security on the WiFi NIC and disabled its power save mode.
We used the 6 GHz band and conducted the experiments in an interference-free environment.
The Ethernet \gls{NIC} supports speeds of up to 1 Gbps, while the WiFi \gls{NIC} supports speeds exceeding 1 Gbps. However, it should be noted that the theoretical maximum throughput of the \gls{AP}'s data plane between the two \glspl{NIC} is around 949 Mbps due to the overhead of networking protocols.

\begin{figure}[!t]
        \centering
        \includegraphics[width=1\linewidth]{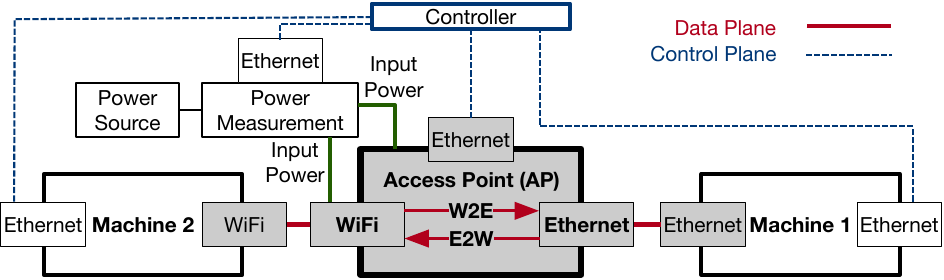}
        \caption{The components, data plane, and control plane of the testbed. Data plane is used for data flow transmissions, and control plane is used for automation of experiments and collecting performance monitoring data.
        }
        \label{fig:testbed}
\end{figure}

Figure \ref{fig:testbed} shows the testbed's components and their connectivity.
The \gls{AP} is the device under test, which is based on the ARM platform mentioned above and runs Linux kernel version 6.7. 
The Ethernet \gls{NIC} of the \gls{AP} is connected to Machine 1, and the WiFi \gls{NIC} of the \gls{AP} is connected to Machine 2.
The \qdisc{} used with the Ethernet interface is FQ-CoDel.
These connections compose the data plane of the testbed.
MTU-size, UDP data flows are generated by Machine 1 or Machine 2, depending on the packet switching path.
For each packet switching path, we utilize UDP instead of TCP to study the impact of packet transmission and reception individually.
Machine 1 and Machine 2 belong to different subnets; therefore, the \gls{AP} performs layer-3 switching.
We disabled \gls{GRO} as its efficiency highly depends on the number of flows and their traffic patterns \cite{cai2021understanding}.
To build a control plane for automating the experiments and performing data collection from various components of the testbed, a machine called Controller is connected to the \gls{AP}, Machine 1, Machine 2, and the power measurement tool.
We added a USB Ethernet dongle to the AP to provide control plane connectivity.
Using this dongle introduces a negligible (less than 7\%) processing overhead on one core.
We developed a control plane daemon to run on the \gls{AP} and collect various types of performance data.
The collected data are sent to the Controller every second.
This daemon is statically assigned to Core 3 of the AP to avoid interfering with packet switching tasks, which are assigned to other cores.

\begin{figure}
    \centering
    \includegraphics[width=0.7\linewidth]{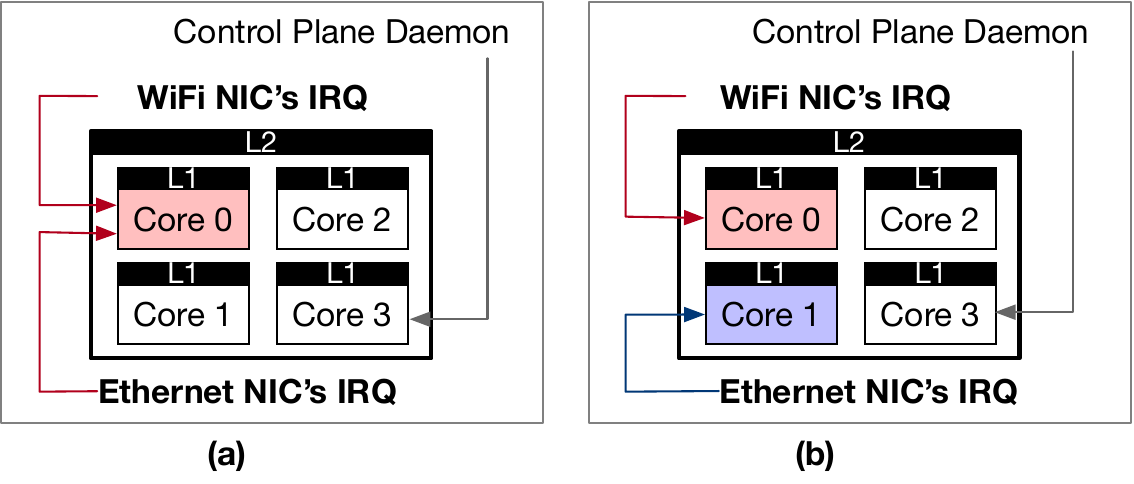}
    \caption{
    (a) Single-core configuration: The \glspl{IRQ} of both \glspl{NIC} are assigned to Core 0.
    (b) Dual-core configuration: The \gls{IRQ} of the WiFi \gls{NIC} is assigned to Core 0 and the \gls{IRQ} of Ethernet \gls{NIC} is assigned to Core 1.
    }
    \label{fig:irq_to_core_rpi}
\end{figure}

Since the \gls{AP}'s processor includes four cores, we investigate the impact of various core assignment configurations on packet switching. To facilitate this, we adjusted the \gls{IRQ} affinity for the Ethernet and WiFi \glspl{NIC}. These configurations, referred to as `single-core' and `dual-core', are depicted in Figure \ref{fig:irq_to_core_rpi}.
It is worth noting that on the \gls{AP} platform, the system enforces that the WiFi \gls{NIC}'s \gls{IRQ} must be assigned exclusively to Core 0, while the Ethernet \gls{NIC}'s \gls{IRQ} can be assigned to any core. 
Therefore, for the single-core configuration, Core 0 is the only feasible option.

We collect various types of performance and operational data from the \gls{AP}.
Linux provides a suite of performance evaluation statistics accessible through the \verb|proc| file system.
The \verb|/proc/interrupts| file offers comprehensive details about each IRQ, including the frequency of \gls{IRQ} arrivals.
The \verb|/proc/softirqs| file represents the number of times RX SoftIRQs and TX SoftIRQs have been invoked.
We utilize \ethtool{} to collect statistics such as the number of frames received and sent. We measure processor utilization using \mpstat{} and collect the number of cycles consumed per core using the \perf{} utility.
The power consumption of the \gls{AP} is measured using a programmable power monitoring tool capable of sampling voltage and current at 1000 samples per second \cite{dezfouli2018empiot}. 
To specifically analyze the power consumption of the \gls{AP}'s processor, we subtract the power consumption of the WiFi \gls{NIC} from the power measurement results.

\section{Performance Analysis and Demystifying Statistics: A Macroalaysis Approach}
\label{macro_analysis}

In this section, we first examine the operation and performance of packet switching in the \glsfirst{E2W} path, followed by the \glsfirst{W2E} path. Subsequently, we compare the differences between the two paths. Specifically, we focus on statistics collected from the \verb|proc| file system, as well as processor utilization, processor cycles, and power consumption.

\subsection{Ethernet-to-WiFi (E2W) Packet Switching}
\label{macro_E2W}

\begin{figure}[!t]
    \centering
    \includegraphics[width=1\linewidth]{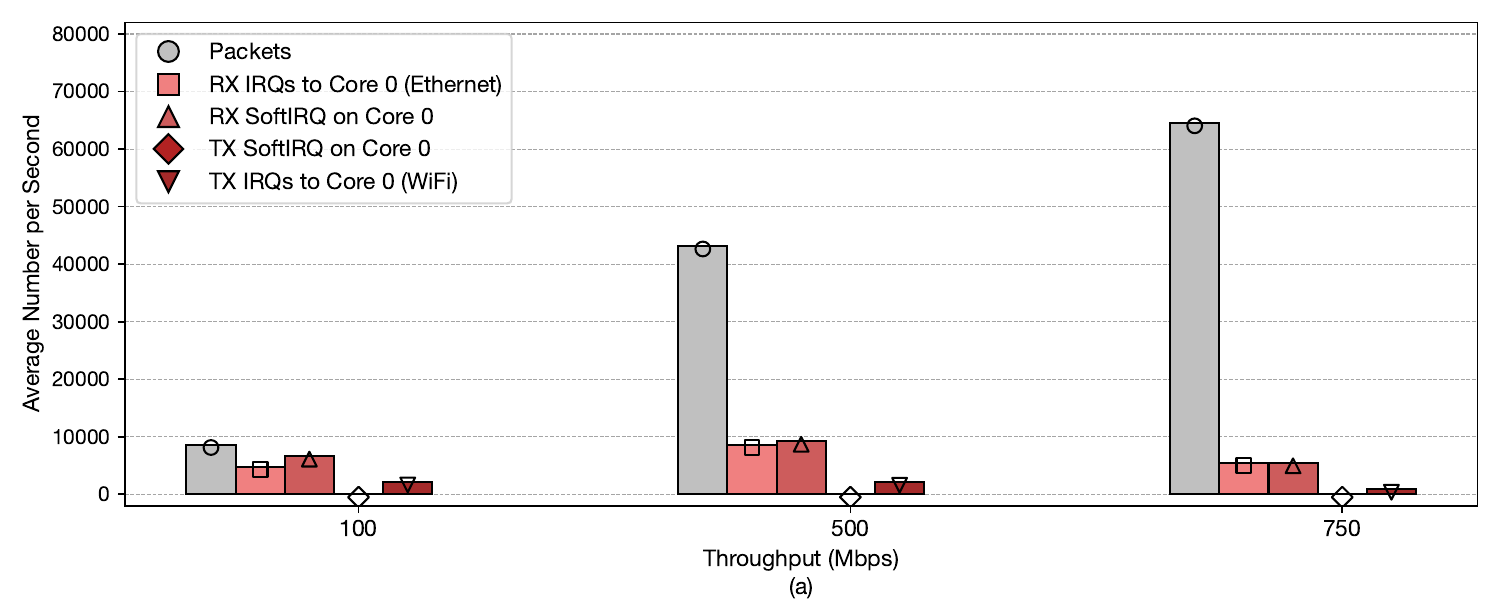}
    \includegraphics[width=1\linewidth]{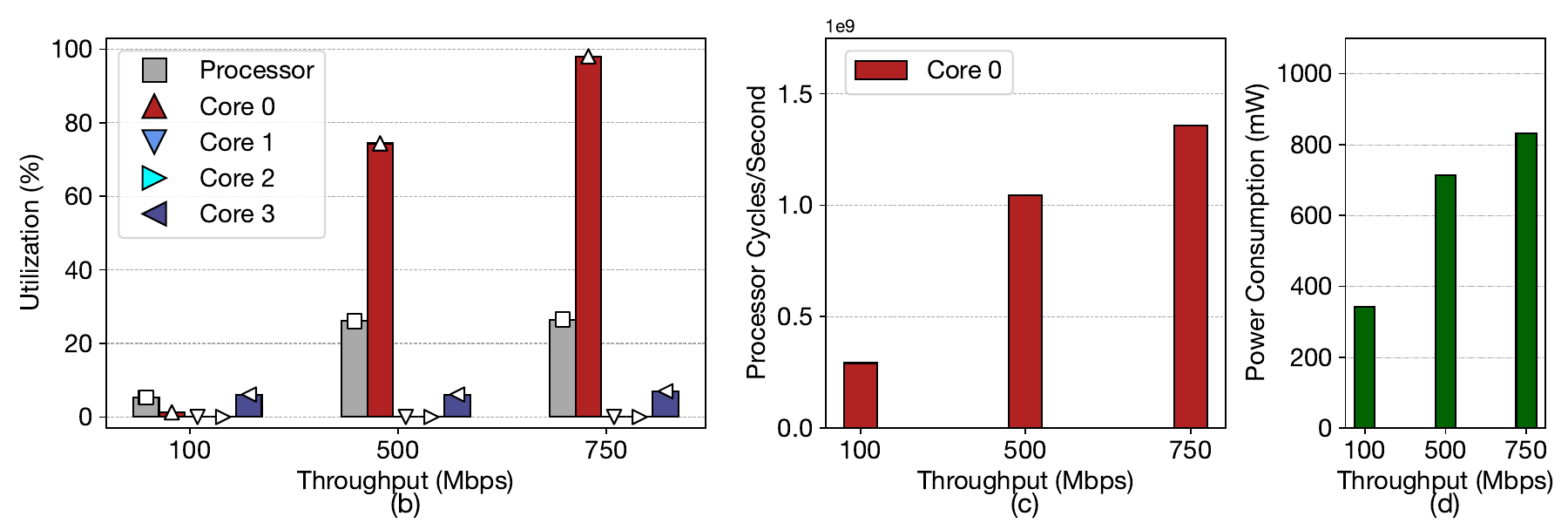}
    \caption{Ethernet-to-WiFi (E2W) packet switching using the single-core configuration. The maximum achieved throughput of this configuration is 750 Mbps.
    }
    \label{fig:wired2wireless_SC}
\end{figure}

\begin{figure}
    \centering
    \includegraphics[width=1\linewidth]{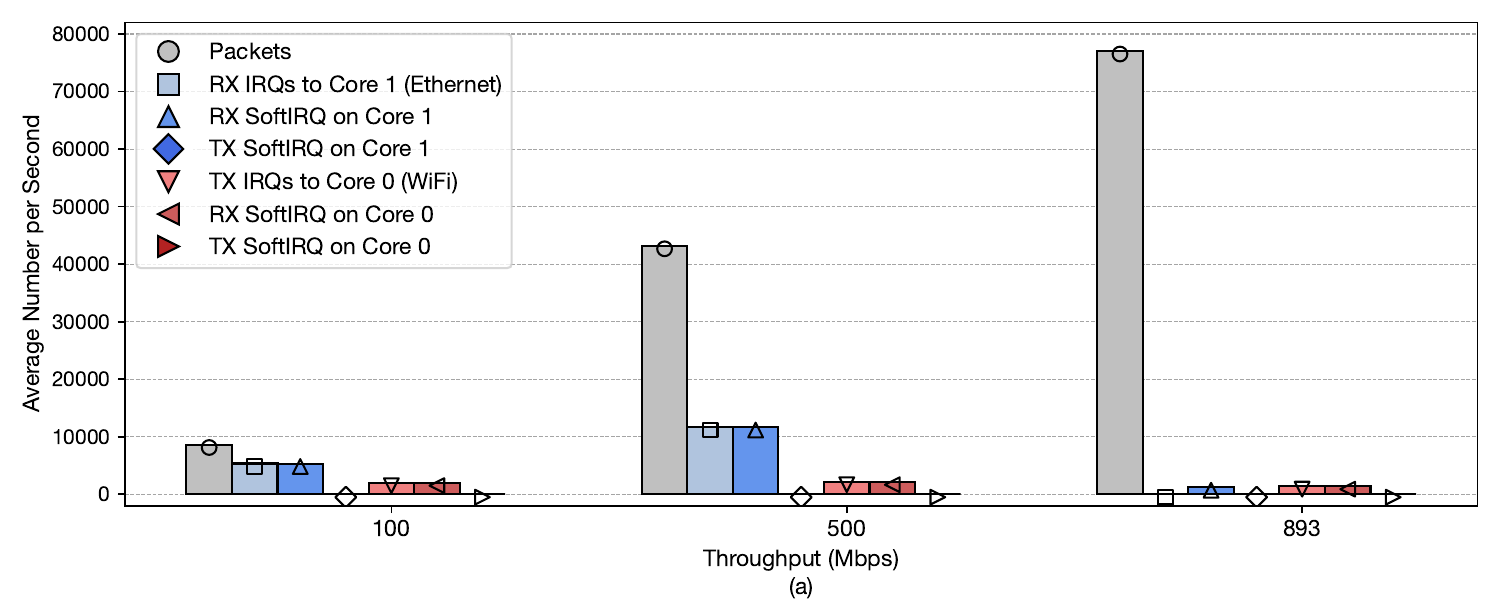}
    \includegraphics[width=1\linewidth]{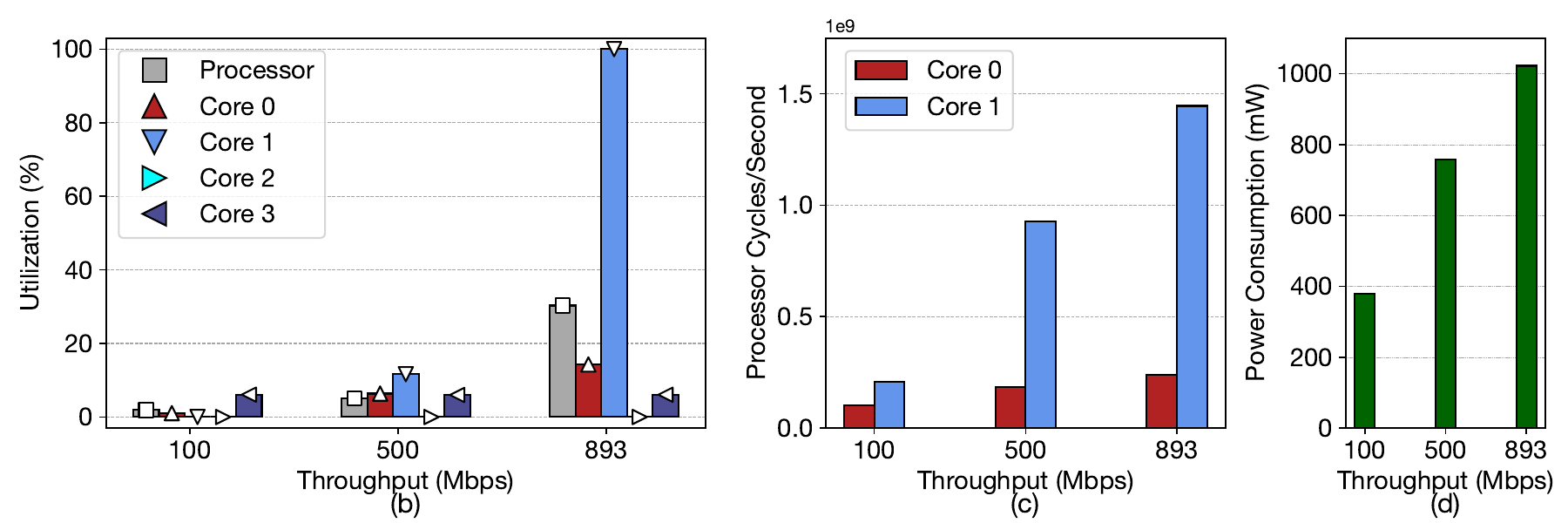}
    \caption{Ethernet-to-WiFi (E2W) packet switching using the dual-core configuration. The maximum achieved throughput of this configuration is 893 Mbps.}
    \label{fig:wired2wireless_DC}
\end{figure}

In this section, we present and discuss the results of \gls{E2W} packet switching on the \gls{AP}, where a UDP flow's packets are received via the Ethernet \gls{NIC} and transmitted by the WiFi \gls{NIC}. 
The results for single-core and dual-core configurations are presented in Figures \ref{fig:wired2wireless_SC} and \ref{fig:wired2wireless_DC}, respectively.
Sub-figures (a) in Figures \ref{fig:wired2wireless_SC} and \ref{fig:wired2wireless_DC} present packet switching statistics.
Sub-figures (b), (c), and (d) demonstrate the utilization of the cores, the number of cycles per core per second, and power consumption, respectively.
In sub-figures (b), `Processor' refers to the average processor utilization across all the cores.
In all these sub-figures, the results are presented for three distinct throughput levels.
We selected two preset throughput levels, specifically 100 and 500 Mbps, alongside the maximum throughput achievable by each configuration.
Notably, the highest supported throughput level for the single-core (Figure \ref{fig:wired2wireless_SC}) and dual-core (Figure \ref{fig:wired2wireless_DC}) configurations are 750 and 893 Mbps, respectively.

\begin{observation}
\label{obsv:rx_irq_vs_sirq}
The number of RX \glspl{IRQ} and RX SoftIRQs are not directly related to the rate of incoming packet processing.
\end{observation}

In Figures \ref{fig:wired2wireless_SC}(a) and \ref{fig:wired2wireless_DC}(a), as the throughput increases from 100 to 500 Mbps, and further to the peak throughput, we observe that the number of Ethernet RX \glspl{IRQ} (denoted as square) rises and then begins to decline.
Remarkably, this count drops to a near-zero value only for the dual-core configuration, as demonstrated in Figure \ref{fig:wired2wireless_DC}(a).

With the dual-core configuration, a maximum throughput of around 893 Mbps is achieved, as shown in Figure \ref{fig:wired2wireless_DC}(a). In this case, the number of Ethernet RX \glspl{IRQ} is approximately two per second, and the number of RX SoftIRQs on Core 1 is about 1140.
It is important to note that the count of RX SoftIRQs (in the \verb|proc| file system) does not equate to the total number of packets processed. This is because each invocation of this SoftIRQ can process multiple packets as long as packets are present in the driver's RX buffer, until the processing quota is exhausted.
To determine the number of packets processed per RX SoftIRQ invocation, we divide the number of packets processed per second by the number of RX SoftIRQs per second, resulting in $73000/1140 \approx 63.8$. This value is very close to the default \gls{NAPI} \verb|weight| of 64 packets (cf. Section \ref{sec:reviews_rx_side}), leading us to conclude that each iteration of the \gls{NAPI} poll processes the maximum number of packets and reschedules repolling of the RX buffer without re-enabling the Ethernet IRQ.
Occasionally, when the number of packets processed falls below 64, the \gls{NAPI} poll concludes and the Ethernet's \gls{IRQ} is re-enabled, resulting in approximately two \glspl{IRQ} per second. The near 100\% utilization of Core 1, as presented in Figure \ref{fig:wired2wireless_DC}(b), confirms that this core is almost fully utilized by the \gls{NAPI} polling mechanism.

\begin{figure}
    \centering
    \includegraphics[width=1\linewidth]{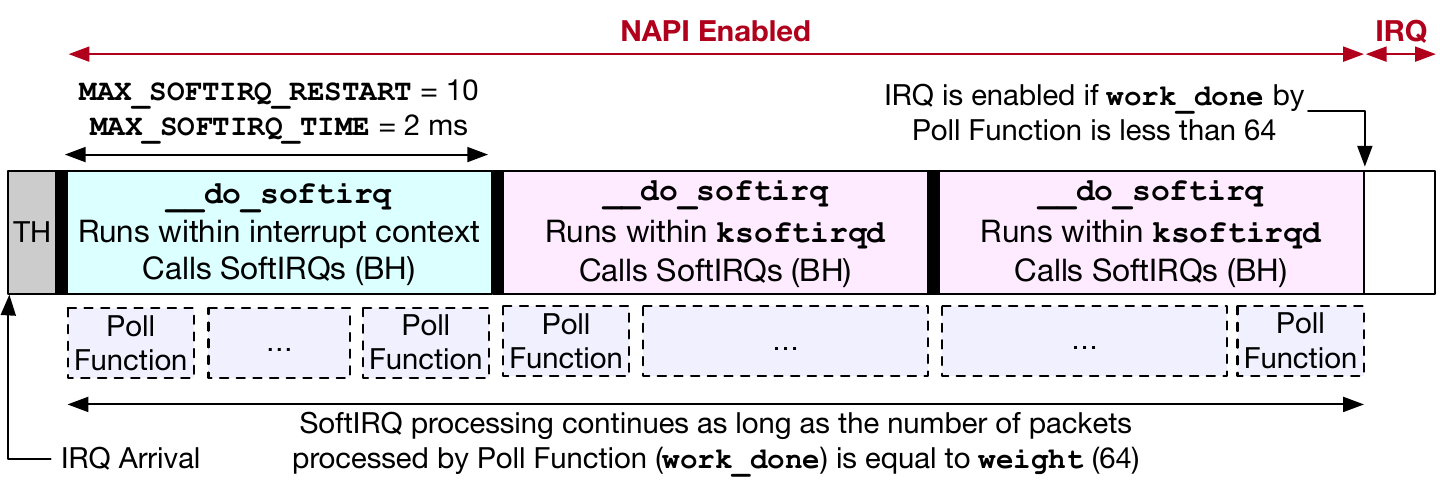}
    \caption{Switching between \gls{NAPI} and \gls{IRQ}. 
    \gls{NAPI} continues as long as the poll function fully utilizes its packet processing budget.
    }
    \label{fig:napi_irq_diagram}
\end{figure}

To better understand the switching between \gls{NAPI} and \gls{IRQ}, we present Figure \ref{fig:napi_irq_diagram} and elaborate on the discussions previously presented in Section \ref{sec:reviews_rx_side}.
When an IRQ's \gls{BH} is triggered, the initial processing is handled by immediate RX SoftIRQ processing within the \verb|__do_softirq| function. This immediate processing, \textit{which runs within the interrupt context}, is essential for rapidly and efficiently handling incoming network traffic. The \verb|__do_softirq| function operates under two main constraints. 
The first constraint, defined by the variable \verb|MAX_SOFTIRQ_RESTART|, is the maximum number of consecutive times the SoftIRQ handling loop can execute the driver's poll function without yielding control back to the system. The second constraint, defined by the variable \verb|MAX_SOFTIRQ_TIME|, specifies the maximum duration the kernel can spend processing SoftIRQs in a single invocation of \verb|__do_softirq|.
In other words, the \verb|__do_softirq| function can invoke the driver's poll function up to 10 times, with each invocation capable of processing up to \verb|weight| packets. Additionally, the total processing duration of these 10 invocations is limited by \verb|MAX_SOFTIRQ_TIME|.
When the packet switching load is high, not all packets can be processed within the interrupt context while enforcing the above constraints. 
Packets exceeding these immediate processing constraints are deferred to \verb|ksoftirqd|, a \textit{kernel thread} dedicated to each core for handling SoftIRQ tasks beyond the immediate \gls{IRQ} context. 
During \verb|ksoftirqd| operation, as long as 64 packets are processed per poll function invocation, the \gls{NAPI} poll is rescheduled, and the \gls{NIC} \gls{IRQ} remains disabled.
Thus, the high utilization of Core 1 in Figure \ref{fig:wired2wireless_DC}(b) is caused by continuously renewing the \gls{NAPI} and running the poll function within the \verb|ksoftirqd| context.

\begin{observation} 
\label{obsr:tx_irq}
The number of RX SoftIRQs is influenced not only by the processing of incoming packets but also by the number of TX IRQs.
More specifically, the egress side also requires RX SoftIRQs, while TX SoftIRQs are not always necessary.
\end{observation} 

For the dual-core configuration, we observe in Figure \ref{fig:wired2wireless_DC}(a) for 100 and 500 Mbps throughput levels that the number of RX SoftIRQs is the same as the number of RX \glspl{IRQ} on Core 1, confirming the activation of an RX SoftIRQ only when an Ethernet RX \gls{IRQ} is triggered.
However, for the single-core configuration, an interesting observation in Figure \ref{fig:wired2wireless_SC}(a) is that the number of RX SoftIRQs is higher than the number of Ethernet RX IRQs, which is particularly evident for the 100 Mbps rate.
This behavior is surprising because when the throughput is low, the arrival of each Ethernet \gls{IRQ} invokes an RX SoftIRQ, which can completely process the packets in the driver's RX buffer and exit without needing to renew the \gls{NAPI} poll. Therefore, we expect the number of RX SoftIRQs to be the same as that of Ethernet RX IRQs.

To rationalize this behavior, we first note that there is one \verb|softnet_data| structure per core (cf. Section \ref{sec:reviews_rx_side}); therefore, for the single-core configuration, the \glspl{IRQ} received from both \glspl{NIC} contribute to the increase in the number of SoftIRQs on that core.
Additionally, by reviewing the Linux source code and using the \ftrace{} utility \cite{ftrace_utility},\textit{ we notice that each TX \gls{IRQ} generated by the WiFi \gls{NIC} (following the transmission of one or more packets) also triggers the activation of a RX SoftIRQ on Core 0.}
In this case, when a TX \gls{IRQ} is generated by the WiFi \gls{NIC}, the \gls{IRQ} handler performs two operations: it adds the driver's poll function to the list of \gls{NAPI} polls for that core and then raises an RX SoftIRQ to call the driver's poll function.
The driver's poll function is then responsible for performing transmission-completion tasks such as TX buffer reclaiming.

\begin{figure}
    \centering
    \includegraphics[width=1\linewidth]{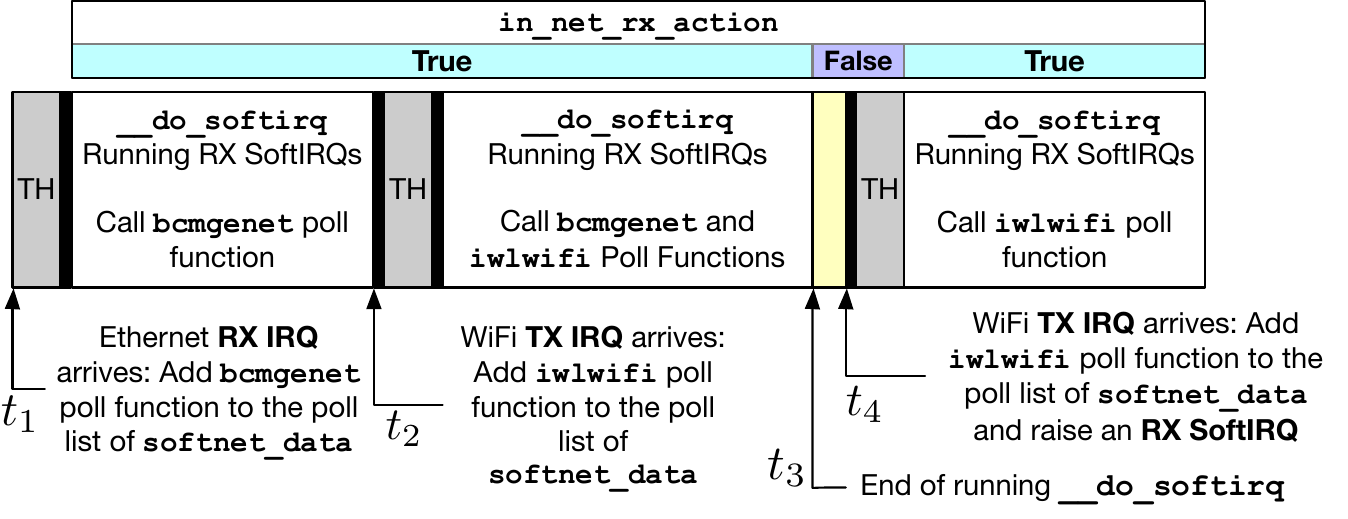}
    \caption{At time instances \(t_2\) and \(t_4\), the WiFi \gls{NIC} generates TX \glspl{IRQ}. In both cases, after adding the poll function of \texttt{iwlwifi} to the poll list of \texttt{softnet\_data}, the RX SoftIRQ is used to call this poll function, which decides if reception, transmission, or both types of packet handling must be performed.
    }
    \label{fig:TX_IRQ_processing}
\end{figure}

To visualize and better understand why RX SoftIRQs are triggered by the egress \gls{NIC}, we analyze the code section that determines which SoftIRQ type must be scheduled.
We explain the kernel's operation by presenting a single-core scenario in Figure \ref{fig:TX_IRQ_processing}.
At time $t_1$, an Ethernet RX \gls{IRQ} arrives, the \gls{TH} processing of this \gls{IRQ} uses the function \verb|__napi_schedule| to add the poll function of the driver to the poll list of the \verb|softnet_data| structure of the core.
When the \verb|__do_softirq| activates an RX SoftIRQ to call the poll function, the \verb|in_net_rx_action| variable is set to \verb|true|.
At time $t_2$, while an RX SoftIRQ is running the poll function of \bcmgenet{} driver, the WiFi \gls{NIC} (egress) generates a TX \gls{IRQ}. 
Again, the function \verb|__napi_schedule| is used to add the poll function of the WiFi \gls{NIC} to the \verb|softnet_data| structure of this core.
At this time, since an RX SoftIRQ is in progress, the \verb|in_net_rx_action| variable of the \verb|softnet_data| structure on this core is \verb|true|; thereby, there is no need to schedule a new SoftIRQ instance.
In other words, the RX SoftIRQ instance that is already running will call the poll function of the \iwlwifi{} driver.

At time $t_3$, there is no ongoing RX SoftIRQ and the value of \verb|in_net_rx_action| is \verb|false|.
When the WiFi \gls{NIC} generates an \gls{IRQ} at time $t_4$, within the \gls{TH} processing, the \verb|__napi_schedule| function calls function \verb|__raise_softirq_irqoff|  (cf. Figure \ref{fig:pkt_switching_steps}) to mark an RX SoftIRQ as pending, ensuring that it will be processed at the next opportunity when the kernel checks for pending SoftIRQs.
Notably, when calling function \verb|__raise_softirq_irqoff| by \verb|__napi_schedule|, the kernel always uses RX SoftIRQ type as the argument of the function.
Therefore, considering the zero number of TX SoftIRQs in Figures \ref{fig:wired2wireless_SC}(a) and \ref{fig:wired2wireless_DC}(a), this phenomenon occurs because TX \glspl{IRQ} are handled by RX SoftIRQs.

\begin{observation}
\label{obsv:proc_util_problem_E2W}
The Linux kernel may not correctly account for the processor cycles consumed by SoftIRQs, depending on the implementation of \gls{IRQ} handler.
We observed this misaccounting problem for the Ethernet driver.
\end{observation}

The results presented in sub-figures (b), (c), and (d) of Figures \ref{fig:wired2wireless_SC} and \ref{fig:wired2wireless_DC} indicate that the number of core cycles and power consumption exhibit similar increasing trends; however, processor utilization does not follow the same pattern.
For the single-core configuration, increasing throughput from 100 to 500 Mbps results in a 71\% jump in utilization of Core 0, and for the dual-core configuration, increasing the throughput from 500 Mbps to the maximum value results in an 8\% increase for Core 0 and 88\% for Core 1.
Another related observation can be made by comparing the actual values of processor utilization and the number of cycles between the single-core and dual-core configurations.
Comparing Figures \ref{fig:wired2wireless_SC}(b) and (c) with Figures \ref{fig:wired2wireless_DC}(b) and (c) for the 500 Mbps throughput, we observe that while the total number of cycles (consumed by Core 0 and Core 1) for the dual-core configuration is higher than that of the single-core configuration (Core 0), the processor utilization of the single-core configuration is higher.

To summarize the above remarks more formally, in Table \ref{tab:normalized_utilization_e2w_sc_dc} we present the \textit{normalized core utilization} and \textit{normalized cycles} consumed by the cores. 
{Normalized core utilization} is calculated by dividing the sum of the utilization of the cores involved in packet switching by the throughput level.
For instance, for the dual-core configuration, since the utilization values of Core 0 and Core 1 for the 500 Mbps throughput are 6.38\% and 11.63\%, respectively, the table reports $(6.38+11.63)/500 = 0.0360$ for this core's normalized utilization per 1 Mbps packet switching rate for the 500 Mbps throughput level.
{Normalized cycles consumed} is calculated by dividing the sum of the percentage of the cycles consumed by the cores involved in packet switching by throughput level.
Here, the percentage of cycles per core is calculated by dividing the number of consumed cycles by the core's frequency, which is 1.5 GHz per core.
For instance, for the dual-core configuration, since the cycles consumed by Core 0 and Core 1 for the 500 Mbps throughput are 182747814 and 926181648, respectively, the table reports $((182747814+926181648)/(1.5\times10^9))\times 100/500 = 0.1479$ for this core's normalized cycles.
The highlighted cells of Table \ref{tab:normalized_utilization_e2w_sc_dc} show the normalized utilization numbers that are much smaller than the normalized utilization values at high throughput levels.
As these cells demonstrate, \textit{the system does not report the correct core utilization values for these throughput levels.}

\begin{table}[ht!]
\centering
\caption{Normalized core utilization and normalized cycles of packet switching cores}
\setlength{\tabcolsep}{3pt} 
\begin{tabular}{l|c|c|c}
\multicolumn{4}{c}{\gls{E2W} with the Single-Core Configuration}
\\
 & {100 Mbps} & {500 Mbps} & {750 Mbps} \\ \midrule
{Core 0 Utilization (\%)}& \cellcolor{gray!20}0.0127 &  0.1488 & 0.1307 \\
{Core 0 Cycles} & 0.1945 &  0.1392 &  0.1206 \\
\midrule
\multicolumn{4}{c}{\gls{E2W} with the Dual-Core Configuration} \\
 & {100 Mbps} & {500 Mbps} & {893 Mbps}\\ \midrule
{Core 0 and 1 Utilization (\%)} & \cellcolor{gray!20} 0.0105 & \cellcolor{gray!20} 0.0360 & 0.1280 \\
{Core 0 and 1 Cycles} & 0.2067 &  0.1479 &  0.1259 \\
\bottomrule
\end{tabular}
\label{tab:normalized_utilization_e2w_sc_dc}
\end{table}

To understand the underlying cause of incorrect processor utilization reporting, we \textit{found that \gls{BH} packet processing may not be correctly accounted for}.
We elaborate on this finding as follows.
In Section \ref{sec:reviews_rx_side} and the discussions of Observation \ref{obsv:rx_irq_vs_sirq}, we explained that after running a \gls{TH}, the kernel invokes a RX SoftIRQ to handle pending \gls{NAPI} functions within the \gls{IRQ} context.
When the restart limit or runtime limit of \verb|__do_softirq| is exhausted, a \verb|ksoftirqd| kernel thread is spawned to handle the remaining RX SoftIRQs.
However, for the Ethernet interface, the RX SoftIRQ instances handled closely following the \gls{TH} and before running a \verb|ksoftirqd| thread are considered part of the `idle' processor utilization, instead of SoftIRQ utilization.
We utilized two approaches to verify this \textit{misaccounting} of SoftIRQ processor cycles.
Firstly, using the \perf{} and \ftrace{} tools, we noticed that the functions of such RX SoftIRQs were recorded under the \verb|idle| or \verb|swapper| threads, and these cycles are added to the `idle' category in \verb|/proc/stat|. 
Therefore, tools such as \mpstat{} misrepresent the actual processor utilization consumed by packet processing.
Secondly, we modified the Linux kernel and changed the \verb|MAX_SOFTIRQ_RESTART| value in the \verb|__do_softirq| function from 10 to 1. 
This change decreased the number of times this function can rerun poll functions without invoking a \verb|ksoftirqd| thread.
Without this change, the processor utilization of Core 1 for the dual-core configuration was around 11\% at 500 Mbps throughput, as Figure \ref{fig:wired2wireless_DC}(b) shows. After applying the change to the kernel, the reported utilization on the same core increased to 67\% (results are not shown in this paper).

It is worth noting that \textit{the misaccounting problem reported above does not occur when processing WiFi-related RX SoftIRQs}. 
This is because, when an Ethernet RX \gls{IRQ} occurs, its immediate RX SoftIRQs are processed as an exception running within the \verb|idle| or \verb|swapper| context.
In contrast, the WiFi-related immediate RX SoftIRQs are processed in the context of threaded IRQs.
We will further explain \textit{threaded IRQs} in Observation \ref{obsv:init_rx_irq_cycles}.

By revealing this misaccounting problem, we can now justify the inaccuracy of the highlighted numbers reported in Table \ref{tab:normalized_utilization_e2w_sc_dc}. 
As throughput increases, the percentage of instances where Ethernet RX SoftIRQs run within the \verb|ksoftirqd| context also increases, leading to more accurate accounting of processor cycles. Consequently, the misaccounting problem diminishes as throughput reaches its maximum value.
For the dual-core configuration, as shown in Table \ref{tab:normalized_utilization_e2w_sc_dc}, the misaccounting problem persists at 100 and 500 Mbps throughput levels. This occurs because Core 1 processes an RX SoftIRQ per RX IRQ, as depicted in Figure \ref{fig:wired2wireless_DC}(a). Therefore, RX SoftIRQs primarily run within the context of the \gls{IRQ} handler, namely, the \verb|idle| or \verb|swapper| contexts, rather than \verb|ksoftirqd|.
In comparison, the single-core configuration exhibits the misaccounting problem only at the 100 Mbps throughput level. In this setup, the percentage of RX SoftIRQs that need to run in the \verb|ksoftirqd| context grows more rapidly than in the dual-core configuration.
Additionally, we identify a secondary, more nuanced reason for this behavior. 
In the single-core configuration, each core has its own \verb|softnet_data| structure (cf. Section \ref{sec:reviews_rx_side}), which both NICs utilize. 
While the WiFi driver is reclaiming its TX buffer entries, an Ethernet RX \gls{IRQ} may be generated, resulting in the addition of a \gls{NAPI} instance for the Ethernet interface to the poll list of the same core's \verb|softnet_data| structure.
Within the \verb|__do_softirq| function, after processing the WiFi driver's function, if the poll list of Core 0 is not empty, the core proceeds with running the poll function of the Ethernet driver within the context of the thread created to handle the WiFi IRQ. 
Therefore, when the \gls{BH} of the WiFi driver is running within a threaded interrupt context, it can execute Ethernet RX SoftIRQs. In this condition, the processor cycles consumed by the Ethernet RX SoftIRQs are properly accounted for under the SoftIRQ processing category.
Therefore, the misaccounting problem related to a \gls{NIC} depends on its driver type, its packet arrival pattern, and the type of other \glspl{NIC} and their associated traffic patterns.

\subsection{WiFi-to-Ethernet (W2E) Packet Switching}
\label{W2E_macro}

\begin{figure}[!t]
    \centering
    \includegraphics[width=1\linewidth]{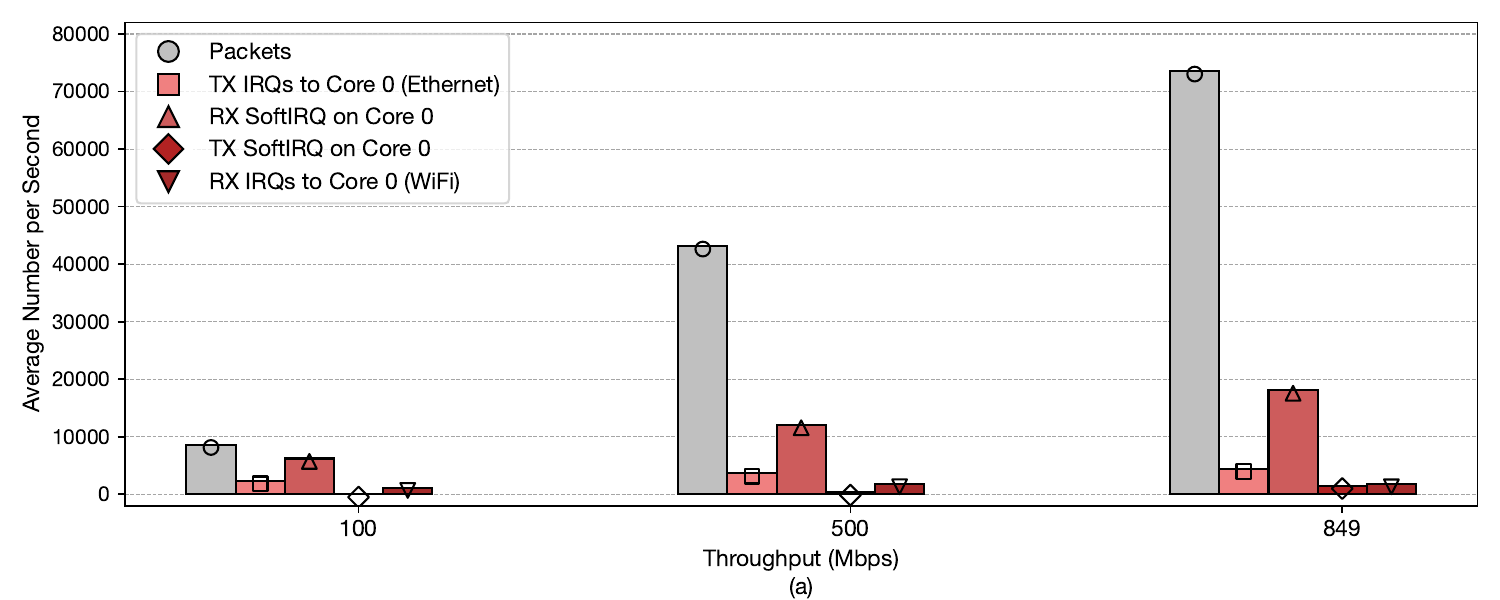}
    \includegraphics[width=1\linewidth]{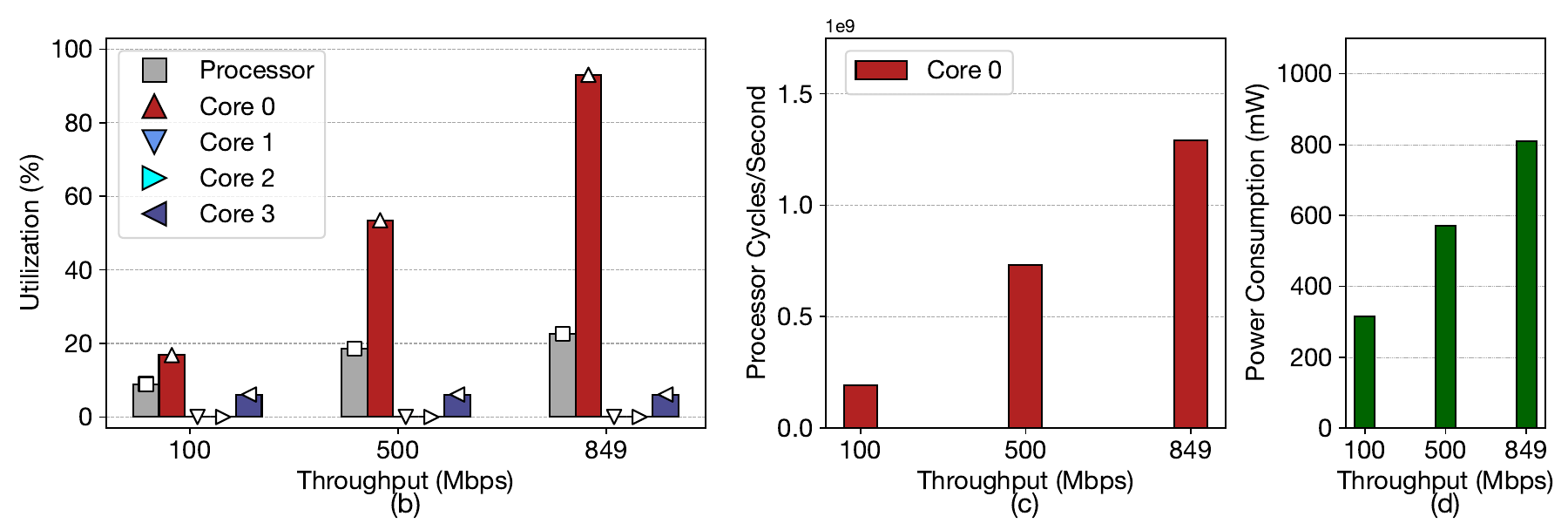}
    \caption{WiFi-to-Ethernet (W2E) packet switching using the single-core configuration.
    The maximum achieved throughput of this configuration is 849 Mbps.
    }
    \label{fig:wireless2wired_SC}
\end{figure}

\begin{figure}
    \centering
    \includegraphics[width=1\linewidth]{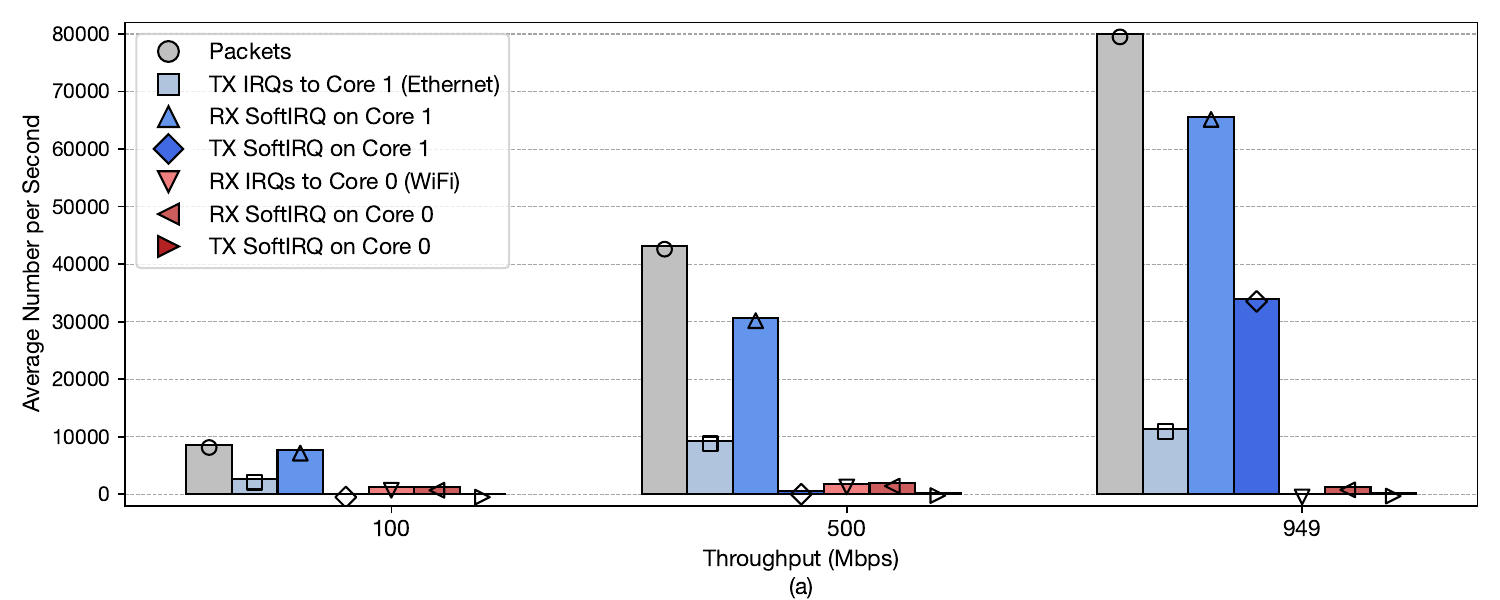}
    \includegraphics[width=1\linewidth]{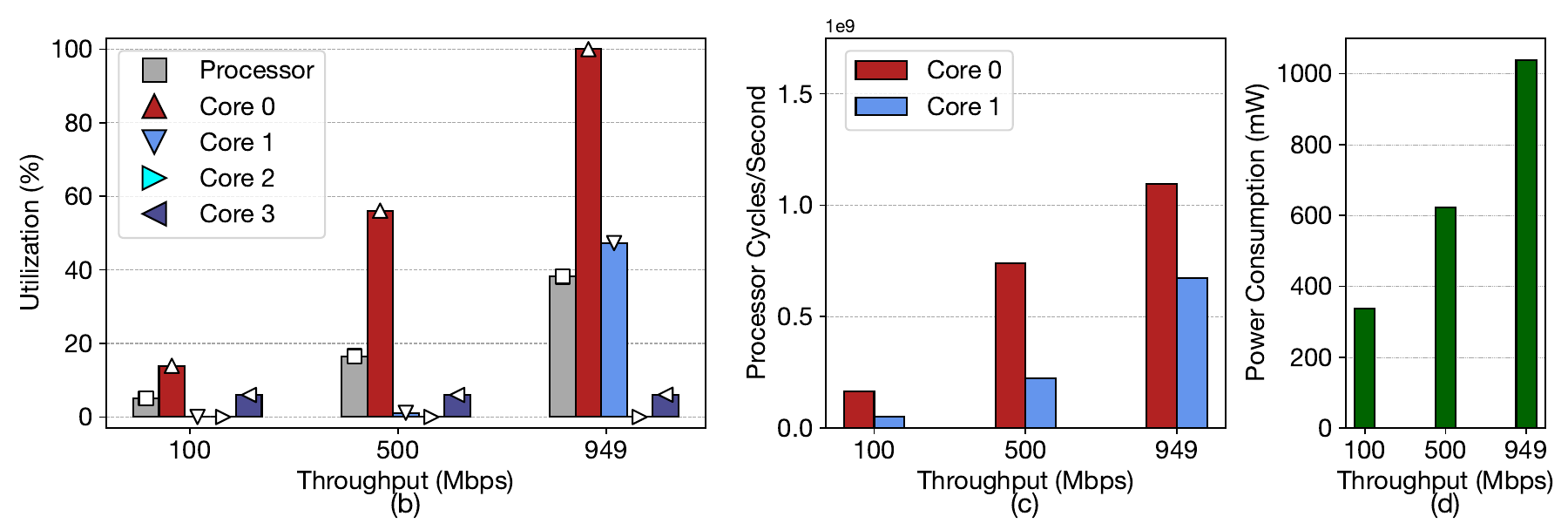}
    \caption{WiFi-to-Ethernet (W2E) packet switching using the dual-core configuration.
   The maximum achieved throughput of this configuration is 949 Mbps. 
    }
    \label{fig:wireless2wired_DC}
\end{figure}

In this section, we present and discuss the results of \gls{W2E} packet switching demonstrated in Figures \ref{fig:wireless2wired_SC} and \ref{fig:wireless2wired_DC}.
The methodologies for collecting and presenting results are the same as those described in Section \ref{macro_E2W}.

\begin{observation}
\label{obsv:tx_side_rx_softirq}
While Linux specifies a protocol for switching between \gls{NAPI} and \gls{IRQ} modes, the \gls{NIC}'s driver may override the protocol.
\end{observation}

In the discussions of Observation \ref{obsr:tx_irq} pertaining to the \gls{E2W} path, we demonstrate that packet transmissions by the WiFi (egress) \gls{NIC} trigger the execution of RX SoftIRQs, which perform tasks such as \gls{TX} buffer reclaiming. 
For the \gls{W2E} path, we observe a similar operation on the Ethernet (egress) side.
The Ethernet \gls{NIC} utilizes the \bcmgenet{} driver. 
When a TX \gls{IRQ} is generated, an RX SoftIRQ is scheduled to call the \verb|bcmgenet_tx_poll| function to reclaim the space used by the transmitted packets.
However, in Figure \ref{fig:wireless2wired_SC}(a), the number of RX SoftIRQs exceeds the sum of Ethernet TX \glspl{IRQ} and WiFi RX \glspl{IRQ}, which is an unexpected observation for low throughput levels such as 100 Mbps.
Additionally, in Figure \ref{fig:wireless2wired_DC}(a), the number of RX SoftIRQs on Core 1 exceeds the number of TX \glspl{IRQ} on this core, which is unexpected for low throughput levels such as 100 Mbps.
For example, for dual-core configuration at 100 Mbps, the average number of packets processed per RX SoftIRQ on the Ethernet (egress) side is $7651/2609 = 2.93$, which is much lower than the default \verb|weight| value of 64 required to reschedule the RX SoftIRQ using the \gls{NAPI} functionality.
We analyze the operations of the \bcmgenet{} driver's \verb|bcmgenet_tx_poll| function to investigate this behavior. 
Typically, a poll function is expected to return the actual number of successfully sent (processed) packets as the \verb|work_done| value.\footnote{For incoming packet processing, as explained in Section \ref{sec:reviews_rx_side} and in Observation \ref{obsv:rx_irq_vs_sirq}, the poll function returns the number of packets processed from the driver's RX buffer.} 
However, the \bcmgenet{} driver's \verb|bcmgenet_tx_poll| function returns the kernel's default \verb|weight| value of 64 as the \verb|work_done|, as long as the actual \verb|work_done| value is greater than 0. 
Thus, if at least one packet buffer is reclaimed, the \gls{NAPI} is renewed to utilize an RX SoftIRQ to execute the \verb|bcmgenet_tx_poll| function again.
We speculate that this \gls{NAPI} renewal method is employed to perform immediate post-transmission actions in anticipation of imminent upcoming transmissions.

\begin{observation}
\textit{The processor cycles misaccounting problem pertaining to the Ethernet driver is also observable when this \gls{NIC} is serving as the egress interface.}
\end{observation}

In the discussions of Observation \ref{obsv:proc_util_problem_E2W}, we explained the processor cycle misaccounting problem when the Ethernet interface is performing ingress packet processing.
The results presented in Figures \ref{fig:wireless2wired_SC} and \ref{fig:wireless2wired_DC} reveal that the problem persists for the \gls{W2E} path as well.
For instance, comparing Figures \ref{fig:wireless2wired_DC}(b) and (c), when the throughput is increased from 500 Mbps to the maximum value, the utilization of Core 1 increases by 4047\% whereas the increase in the number of cycles consumed by this core is 176\%.
Nevertheless, the misaccounting problem is less obvious for the \gls{W2E} path, compared to the \gls{E2W} path.
%

For the dual-core configuration, the main reason for the mitigated misaccounting problem in the \gls{W2E} path is as follows. In the \gls{W2E} path, the ingress packet processing is handled by the WiFi interface, which utilizes threaded \glspl{IRQ}. Threaded \glspl{IRQ} allow for the correct accounting of processor cycles consumed by SoftIRQs that run immediately after \gls{IRQ} handling and before deferring the work to \verb|ksoftirqd|. 
Therefore, for the \gls{W2E} path, only the SoftIRQ instances that handle egress processing and are not executed within the \verb|ksoftirqd| context may be misaccounted.
%
%
For the single-core configuration, in addition to the above reason, there is a secondary cause.
As discussed in Observation \ref{obsv:proc_util_problem_E2W}, the processor cycles consumed by the SoftIRQ instances that run within the context of the threaded \gls{IRQ} handler of the WiFi interface are properly accounted for.
In the \gls{W2E} path, as the number and duration of such threaded \gls{IRQ} handlers increase, so does the SoftIRQ instances of the Ethernet interface that run within those threads.
Therefore, the cycles of more Ethernet SoftIRQs are properly accounted for.

\subsection{Comparison of \gls{E2W} and \gls{W2E}}
\label{macro_compare}

Comparing the results for \gls{W2E} and \gls{E2W} paths, we observe significant differences in the performance and operation of packet switching for these two paths. We identify and justify these observations as follows.

\begin{observation}
\label{obsv:core_utilization_inequalty}
\textit{For the dual-core configuration, the \gls{W2E} path inherently utilizes two cores, whereas the \gls{E2W} path utilizes one core overwhelmingly.}
\end{observation}

By default, each network interface uses a \qdisc{} as its entry point into the \netdev{} subsystem, as discussed in Section \ref{sec:reviews_tx_side}. However, \textit{WiFi interfaces may not utilize a \qdisc{}.} Specifically, to implement airtime fairness for downlink packet delivery, \glspl{AP} employ sophisticated queuing algorithms within the MAC layer of WiFi interfaces \cite{hoiland2017ending}. This part of the MAC layer, which is implemented as the \maceighteleven kernel module, integrates the FQ-CoDel queue management with airtime fairness methods. Consequently, \qdisc{} is completely bypassed on the WiFi interfaces.
Considering this configuration and referring to Figure \ref{fig:pkt_switching_steps}, we explain the operation of \gls{E2W} packet switching. When a packet received at the Ethernet interface reaches the function \verb|__dev_queue_xmit|, it bypasses the \qdisc{} and proceeds to the function \verb|dev_hard_start_xmit|. This function forwards the packet to the \maceighteleven module for queuing and further processing before it is sent to the driver's TX buffer. 
It is important to note that after ingress Ethernet packet processing, the \maceighteleven packet processing continues within the interrupt or \verb|ksoftirqd| context and runs on the core that is handling \glspl{IRQ} for the Ethernet interface; thereby, \textit{packet processing from the ingress \gls{NIC} to the egress \gls{NIC} is handled by a single core, which is Core 1 in our testbed.}
Conversely, in the \gls{W2E} path, a certain percentage of packets are added to the Ethernet \gls{NIC}'s \qdisc{} by the core processing incoming packets from the WiFi interface. Then the core handling the Ethernet \glspl{IRQ} performs \qdisc{} processing and adds the packets to the TX buffer. 
This analysis of packet switching operations explains why the \gls{W2E} path achieves higher throughput compared to \gls{E2W}: \textit{\gls{W2E} inherently utilizes multi-core processing, provided that the \glspl{IRQ} of the two \glspl{NIC} are assigned to different cores.}

The number of TX SoftIRQs is related to the above discussion.
In the results pertaining to \gls{E2W} packet switching demonstrated in Section \ref{macro_E2W}, Figures \ref{fig:wired2wireless_SC}(a) and \ref{fig:wired2wireless_DC}(a) show a value of zero for the number of TX SoftIRQs.
In these results, for the ingress (Ethernet) side, it is reasonable to have a zero number of TX SoftIRQs because the UDP flow is unidirectional and the Ethernet \gls{NIC} is the ingress interface.
However, we expect to see TX SoftIRQs on the egress (WiFi) side.
In contrast, for the \gls{W2E} path, Figures \ref{fig:wireless2wired_SC}(a) and \ref{fig:wireless2wired_DC}(a) show a considerable number of TX SoftIRQs, especially for the maximum throughput levels.
\textit{In the \gls{W2E} path, processing and running the \qdisc{} requires scheduling TX SoftIRQs.}
More specifically, when the \qdisc{}'s quota is used, the \verb|__netif_schedule| function (cf. Figure \ref{fig:pkt_switching_steps}) is called to schedule a TX SoftIRQ, which is used for dequeuing packets from the \qdisc{} by running the \verb|net_tx_action| function. 
In Section \ref{micro_comp_directions}, Observation \ref{obsv:qdisc_distribution}, we will explain how lowering the throughput increases the number of packets that can bypass the \qdisc{}, which in turn reduces the number of TX SoftIRQs.

\begin{observation}
\label{obsv:power_first_E2W}
For both \gls{E2W} and \gls{W2E} paths, the dual-core configuration consumes more power than the single-core configuration at all throughput levels.
The normalized power consumption of the \gls{E2W} path is higher than that for \gls{W2E} across all throughput levels.
\end{observation}

In sub-figure (d) of Figures \ref{fig:wired2wireless_SC}, \ref{fig:wired2wireless_DC}, \ref{fig:wireless2wired_SC}, and \ref{fig:wireless2wired_DC}, we observe the following: 
(i) For both \gls{E2W} and \gls{W2E} paths, the power consumption of the dual-core configuration is higher than that of the single-core configuration, and 
(ii) the power consumption of the \gls{E2W} path is higher than that of \gls{W2E}.
To better present these differences, Table \ref{tab:power_consumption_ratio} summarizes the normalized power consumption computed as milliwatts (mW) per 1 Mbps of throughput for three throughput levels. 
Table \ref{tab:processor_ratio} shows normalized cycles consumed by the packet switching cores.
By comparing these two tables, we observe a clear numerical relationship between processor cycles and power consumption.

\begin{table}[ht!]
\centering
\caption{Normalized power consumption values of the \gls{AP}}
\setlength{\tabcolsep}{1.7pt}
\begin{tabular}{l|c|c|c}
 & {100 Mbps} & {500 Mbps} & {Max}\\ \midrule
\rowcolor{white}{\gls{E2W} Single-Core} (mW) & 3.410 & 1.426 & 1.108 (750 Mbps)\\
\rowcolor{white}{\gls{E2W} Dual-Core} (mW) & 3.780 & 1.514 & 1.144 (893 Mbps)\\
\rowcolor{gray!20}
{\% Dual-Core vs. Single-Core} & 10.85\% & 6.17\% & 3.21\%\\
\midrule
\rowcolor{white}{\gls{W2E} Single-Core} (mW) & 3.140 & 1.142 & 0.953 (850 Mbps)\\
\rowcolor{white}{\gls{W2E} Dual-Core} (mW) & 3.370 & 1.244 & 1.095 (949 Mbps)\\
\rowcolor{gray!20}
{\% Dual-Core vs. Single-Core} & 7.32\% & 8.93\% & 8.60\%\\
\bottomrule
\rowcolor{gray!20}
\% \gls{E2W} vs. \gls{W2E} for Single-Core & 8.60\% & 24.86\% & 16.26\%\\
\rowcolor{gray!20}
\% \gls{E2W} vs. \gls{W2E}  for Dual-Core & 12.16\% & 21.70\% & 10.53\%\\
\bottomrule
\end{tabular}
\label{tab:power_consumption_ratio}
\end{table}

\begin{table}[ht!]
\centering
\caption{Normalized processor cycles of packet switching cores}
\setlength{\tabcolsep}{3pt}
\begin{tabular}{l|c|c|c}
  & {100 Mbps} & {500 Mbps} & {Max}\\ \midrule
\rowcolor{white}{\gls{E2W} Single-Core} & 0.1945 & 0.1392 & 0.1206 (750 Mbps)\\    
\rowcolor{white}{\gls{E2W} Dual-Core} & 0.2067 & 0.1479 & 0.1259 (893 Mbps)\\
\rowcolor{gray!20}{\% Dual-Core vs. Single-Core} & 6.23\% & 6.24\% & 4.40\% \\
\midrule
\rowcolor{white}{\gls{W2E} Single-Core} & 0.1277 & 0.1088 & 0.1011 (850 Mbps)\\
\rowcolor{white}{\gls{W2E} Dual-Core} & 0.1440 & 0.1285 & 0.1242 (949 Mbps)\\
\rowcolor{gray!20}{\% Dual-Core vs. Single-Core} & 12.8\% & 18.1\% & 22.7\% \\
\bottomrule
\rowcolor{gray!20}
\% \gls{E2W}vs.\gls{W2E} of Single-Core & 52.4\% & 27.90\% & 19.21\%\\
\rowcolor{gray!20}
\% \gls{E2W}vs.\gls{W2E} of Dual-Core & 43.56\% & 15.07\% & 1.39\%\\
\bottomrule
\end{tabular}
\label{tab:processor_ratio}
\end{table}

\begin{table}[ht!]
\centering
\caption{Normalized number of cache invalidations for packet switching cores}
\setlength{\tabcolsep}{3pt}

\begin{tabular}{l|c|c|c}
  & {100 Mbps} & {500 Mbps} & {Max}\\ \hline
\rowcolor{white}{\gls{E2W} Single-Core} & 7.16 & 1.46 & 1.04 (750 Mbps)\\
\rowcolor{gray!20}{\gls{E2W} Dual-Core} & 290.2867 & 147.55 & 89.34 (893 Mbps)\\
\midrule
\rowcolor{white}{\gls{W2E} Single-Core} & 6.91 & 1.39 & 0.89 (850 Mbps)\\
\rowcolor{gray!20}{\gls{W2E} Dual-Core} & 1253.5 & 694.14 & 476.65 (949 Mbps)\\
\bottomrule
\end{tabular}
\label{tab:cache_invalidation_ratio}
\end{table}

We identify three key factors that contribute to the higher normalized processor cycles and, consequently, the increased normalized power consumption in the dual-core configuration.
First, when a single core handles all the packet processing tasks from ingress to egress, the core checks the RX buffer less frequently than the case where a dedicated core handles ingress processing. 
Accordingly, as the chance of packet accumulation increases, more packets are processed per RX SoftIRQ, which means the \glspl{IRQ} are enabled and processed less often.
This can be observed in Figures \ref{fig:wired2wireless_SC}, \ref{fig:wired2wireless_DC}, \ref{fig:wireless2wired_SC}, and \ref{fig:wireless2wired_DC} as the number of RX \glspl{IRQ} and RX SoftIRQs are lower for the single-core configuration. 
For instance, on the \gls{E2W} path, the increase in the number of \glspl{IRQ} for the dual-core configuration compared to the single-core configuration is 6\% and 28.8\% for 100 and 500 Mbps throughput levels, respectively, and the number of RX SoftIRQs increase by 10.9\% and 50.4\%, respectively.
Second, we identify that the dual-core configuration results in a higher number of \gls{L1-DCache} invalidations.
Table \ref{tab:cache_invalidation_ratio} shows the normalized number of cache invalidations calculated by dividing the total number of cache invalidations by the throughput level.
The highlighted rows of the table reveal higher cache invalidation rate for the dual-core configurations.
As Figure \ref{fig:irq_to_core_rpi} shows, each core has a dedicated L1 cache; therefore, for the dual-core configuration, the packets in Core 1's L1 cache must transfer to Core 0's L1 cache during the packet switching process, resulting in a higher number of cache invalidation.
It is worth noting that for the \gls{E2W} path, even though Core 1 handles both ingress and egress processing, Core 0 handles TX \glspl{IRQ} for packet buffer claiming; therefore, Core 0 requires access to the SKBs transmitted.
Third, when two cores handle packet-switching processes, the number of context switches between kernel tasks increases. 
For the \gls{E2W} path and 500 Mbps throughput, we measure these values for the single-core and dual-core as 3361 and 5824, respectively.
For the \gls{W2E} path, these numbers are 3351 and 4369, respectively.

While the above mentioned factors contribute to the higher power consumption of dual-core configuration, we leave quantitative analysis of the impact of each factor to power consumption as a future work.
The above factors also justify the lower normalized power consumption and lower normalized processor cycles as throughput increases.
For instance, the decrease in the number of cache invalidations versus throughput can be observed in Table \ref{tab:cache_invalidation_ratio}.

\section{Analyzing the Stages of Packet Switching: A Microanalysis Approach}
\label{micro_comp_directions}

\begin{figure}[!ht]
    \centering
    \includegraphics[width=1\linewidth]{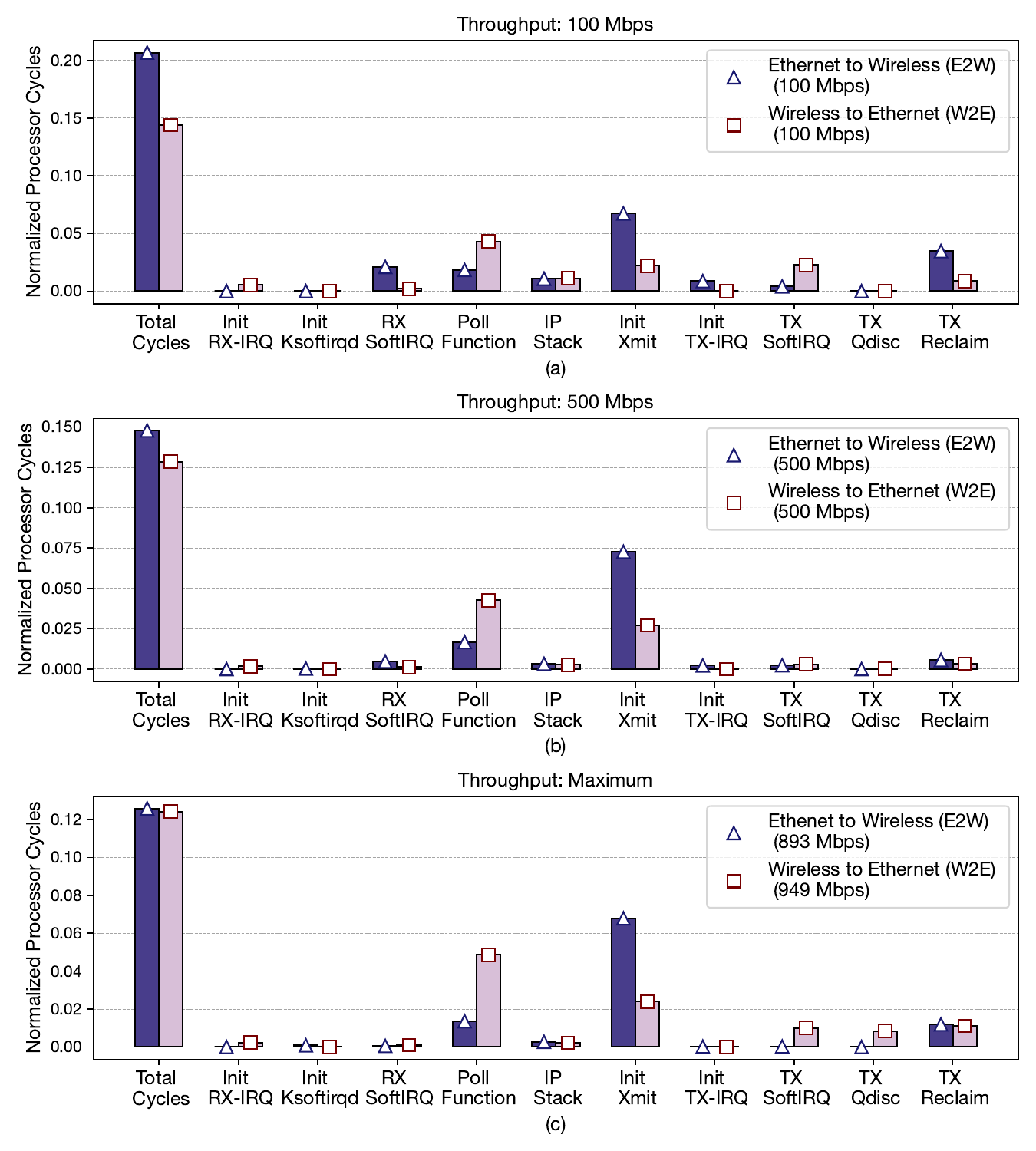}

    \caption{The ``Total Cycles" on the x-axis refers to the normalized cycles consumed by all the cores.
    For each packet switching stage, the value refers to the normalized number of cycles consumed by that stage only.
    The results are for the dual-core configuration for (a) 100 Mbps, (b) 500 Mbps and (c) maximum throughput levels.
    The x-axis corresponds to the various stages outlined in Table \ref{tab:perf_x_axis}. 
   }
    \label{fig:flame_graph}
\end{figure}

The metrics and analyses presented in the previous section enabled us to describe the overarching operational framework and discern the differences in packet switching considering path and core assignment configuration.
To deepen our understanding of these operations and the root causes of the observed differences, this section adopts a more granular approach by conducting a micro-analysis of packet switching stages.
Our methodology is structured as follows.
We decompose packet switching operations into distinct stages, as Table \ref{tab:perf_x_axis} illustrates.
Figure \ref{fig:flame_graph} presents the normalized number of core cycles for the three throughput levels and both \gls{E2W} and \gls{W2E} paths using the dual-core configuration.
We used the same normalization method explained in Observation \ref{obsv:proc_util_problem_E2W}.
To quantify the processing load associated with each stage, we measured the number of processor cycles using the \perf{} tool. 
The details of the mappings and the corresponding codes are available in the GitHub repository of this paper \cite{wifi-ap-github}.
\begin{table}
    \caption{Definition of the stages in packet switching}
    \label{tab:perf_x_axis}
    \centering
    \begin{tabular}{|>{\centering\arraybackslash}m{0.07\textwidth}|m{0.37\textwidth}|}
    \hline
        \textbf{Stage} & \textbf{Description} (cf. Figure \ref{fig:pkt_switching_steps} for  more information) \\
        \hline 
        \hline
        Total Cycles & The number of processor cycles per second consumed by the cores assigned to packet processing.
        This includes packet switching and other kernel operations.
        \\
        \hline
        \hline
        \multicolumn{2}{|c|}{\cellcolor{lightgray!25}Ingress Packet Processing}
        \\
        \hline
        \initRXIRQ{} & 
        The operations between the detection of an RX \gls{IRQ} and the start of running an RX SoftIRQ.
        \\
        \hline
        \initKsoftirqd{} & The operations between the creation of a \verb|ksoftirqd| thread and the start of running a SoftIRQ.
        \\
        \hline
        \RXIRQ{} & 
        The execution of an RX SoftIRQ until the start of running the driver's poll function. 
        This includes the functions \verb|__do_softirq| and \verb|net_rx_action|, excluding the actual execution of driver's poll function.
        \\
        \hline
        \pollFunction{} & 
        The execution of the driver's poll function, which includes fetching packets from the device's RX buffer, performing necessary processing and passing the packets up the IP stack.
        \\
        \hline
        \hline
        \multicolumn{2}{|c|}{\cellcolor{lightgray!25}IP Stack Processing}
        \\
        \hline        
        \ipStack{} & 
        The execution of IP stack for processing incoming packets. This includes the operations starting \verb|ip_rcv| and before \verb|__dev_queue_xmit|.
        \\
        \hline
        \hline
        \multicolumn{2}{|c|}{\cellcolor{lightgray!25}Egress Packet Processing}
        \\
        \hline
        \initXmit{} & 
        The execution of the functions responsible for pushing the packet into the \qdisc{} or the driver's TX buffer.
        This includes the operations started by the \verb|__dev_queue_xmit| function.
        \\
        \hline
        \initTXIRQ{} & The operations between the detection of a TX \gls{IRQ} and the start of running a TX SoftIRQ.
        \\
        \hline
        \TXIRQ{} & 
        The execution of a SoftIRQ until the start of running the driver's poll function to perform the \TXReclaim{} stage or \TXQdisc{} stage.
        This includes the functions \verb|__do_softirq|, \verb|net_rx_action| and \verb|net_tx_action| (cf. Observation \ref{obsr:tx_irq} for more information).
        \\
        \hline
        \TXQdisc{} & 
        The execution of \verb|qdisc_run|, which processes the packets that have been enqueued in the \qdisc{} and pushes them into the driver's \gls{TX} buffer.
        \\
        \hline
        \TXReclaim{} &
        The process of reclaiming transmission resources (SKBs) after packets have been sent out. 
        \\
        \hline
    \end{tabular}
\end{table}

\begin{observation}
\label{obsv:init_rx_irq_cycles}
The difference in the number of processor cycles consumed during the \initRXIRQ{} stage for \gls{E2W} and \gls{W2E} paths is due to the distinct \gls{IRQ} handling methods employed by the WiFi and Ethernet drivers.
\end{observation}

Figure \ref{fig:flame_graph} shows that the normalized number of core cycles consumed by the \initRXIRQ{} stage of the \gls{W2E} path are 0.005, 0.001, and 0.002 for the 100, 500 Mbps, and maximum throughout levels, respectively; whereas, for the \gls{E2W} these numbers are measured as zero.
This difference is caused by the use of \textit{threaded interrupt handler} by the \iwlwifi{} driver, as we briefly mentioned in Observation \ref{obsv:proc_util_problem_E2W}.
Threaded interrupt handlers are designed to improve system responsiveness by allowing complex interrupt handling to be offloaded to kernel threads \cite{rothberg2015interrupt}. 
Some WiFi drivers (e.g., \iwlwifi{} \cite{iwlwifiWiki} and Realtek RTW89 \cite{realtek_rtw89_threadedIRQ}) handle complex tasks like multiple stream management, encryption/decryption, and dynamic signal strength adaptation. These operations can be resource-intensive and may involve blocking, which makes threaded handlers particularly advantageous. 
When an \gls{IRQ} arrives, the primary interrupt handler can quickly acknowledge the interrupt and defer the complex processing to a threaded handler, which runs in a process context and can safely sleep or wait for resources. This approach minimizes latency in the interrupt path and allows for better prioritization and scheduling of tasks.
In contrast, most Ethernet interfaces, with their more straightforward processing requirements, rely on the primary handler to manage interrupt processing efficiently.

With the above discussion, we take a closer look into the operation of the \iwlwifi{} driver. 
In the \gls{W2E} path, when a WiFi \gls{IRQ} arrives, the kernel's interrupt handling code runs \verb|iwl_pcie_msix_isr|, which is the primary interrupt handler for this \gls{NIC}. The primary handler identifies the specific cause of the interrupt using the index associated with the \gls{MSIX} vector.
Based on the identified cause, \verb|iwl_pcie_msix_isr| invokes a secondary function, \verb|iwl_pcie_irq_rx_msix_handler|, as a threaded handler to process the interrupt further.
During the execution of the threaded handler, the local \glspl{BH} on the processor assigned to this \gls{IRQ} are disabled to prevent re-entrant execution while handling critical sections of the threaded handler. 
Within the \gls{TH} execution, once the critical sections are processed and the threaded handler completes, the \gls{TH} must explicitly re-enable the local \glspl{BH} using the \verb|__local_bh_enable_ip| function.
By relying on the \perf{} tool, we observed that most of the processor cycles consumed by the \initRXIRQ{} stage of the \gls{W2E} path are related to the \verb|__local_bh_enable_ip| function. This is because this function needs to check various conditions to ensure it is safe to re-enable preemption and interrupts. 
In contrast to the \iwlwifi{} handler, the Ethernet \gls{IRQ} handler does not require a threaded handler. This makes the Ethernet \gls{IRQ} handler more efficient as it avoids the additional overhead of managing a threaded context and \glspl{BH}.

\begin{observation}
\label{obsv: xmit_poll_analysis}
\textit{Among all the stages, the two most process-intensive stages are the \pollFunction{} stage and the \initXmit{} stage. For the WiFi interface in particular, the processing load of these stages is higher than that of the Ethernet interface. 
}
\end{observation}

The \pollFunction{} stage of the WiFi ingress processing incurs higher overhead than that of the Ethernet interface.
Figure \ref{fig:flame_graph} shows that for 100, 500 Mbps, and maximum throughput levels, the number of cycles consumed by the \pollFunction{} stage in the \gls{W2E} path are 2.3x, 2.5x, and 3.8x higher, respectively, than in the \gls{E2W} path.
This observation is because the poll function of the \iwlwifi{} driver performs more operations than the \bcmgenet{} driver.
After retrieving the incoming packets from the RX buffer, the \bcmgenet{} driver performs necessary validations (e.g., error checking, packet size validation) and passes the packets to the IP stack for further processing.
In contrast, the \iwlwifi{} driver performs more complex tasks and needs to process 802.11 headers, which are more intricate than Ethernet headers. This involves extracting and processing additional fields such as QoS control, sequence control, and security-related fields. After parsing the headers of each \gls{MPDU}, the driver performs validations such as error checking and packet size validation on each \gls{MPDU}. If a \gls{MPDU} is part of a fragmented frame sequence, the driver handles the fragmentation and reassembly of packets. After these parsing and validation steps, the driver strips off the 802.11 headers and additional encapsulation headers before handing the packet to the IP layer.

The \initXmit{} stage of the WiFi egress processing incurs higher overhead than that of the Ethernet interface.
Figure \ref{fig:flame_graph} shows that for the 100, 500 Mbps and maximum throughput levels, the processor cycles consumed by the \initXmit{} stage of the \gls{E2W} path are 3x, 2.7x and 2.6x higher, respectively, than the \initXmit{} stages of the \gls{W2E} path.
Recall that the \initXmit{} stage starts after the completion of IP stack processing.
At this point, the packet is passed to the \netdev{} subsystem of the egress \gls{NIC}. 
For the \gls{E2W} path, since there is no \qdisc{} (cf. Section \ref{macro_analysis}, Observation \ref{obsv:core_utilization_inequalty}), packets are passed to the \maceighteleven module. This module implements complex queuing methods for establishing airtime and QoS-aware delivery of egress traffic. 
More specifically, the \maceighteleven module strives to allocate a fair share of airtime to each station, and the priority of the traffic (i.e., voice, video, best-effort, and background) also affects the order of packet transmissions \cite{eaps_jay,hoiland2017ending}.
These operations introduce queue management and locking overheads that increase with throughput.
Below the \maceighteleven module, the \iwlwifi{} driver implements tasks such as retransmissions, rate control, power management, medium access control, encryption, and QoS. For instance, based on the traffic type, the driver enforces hardware-specific configurations for channel access. All of these factors contribute to the higher processing load of egress processing compared to ingress processing.
Our deeper analysis (not included in this paper) of \gls{E2W} path revealed that the \maceighteleven processing accounts for 62.6\% of the \initXmit{} sub-stage. 

For the \gls{W2E} path, the tasks for the \initXmit{} stage are fewer and less complex. 
Specifically, for those packets that are directly transmitted, the processing load of the \bcmgenet{} driver is less than that of the \maceighteleven module and \iwlwifi{} driver. This is because the \bcmgenet{} driver does not need to handle complex tasks such as QoS-based queuing and channel access. 
However, as the throughput increases, the number of packets that cannot bypass the \qdisc{} increases, and the overhead of the \qdisc{} also rises. 
We elaborate on the processing load of the \qdisc{} as follows.

\begin{observation}
\label{obsv:qdisc_distribution}
The \gls{W2E} path is more efficient than the \gls{E2W} one. 
However, as throughput increases, the difference in processing overhead between these two paths reduces.
\end{observation}

Comparing Figures \ref{fig:flame_graph}(a), (b), and (c), we observe that for all throughput levels, the number of cycles consumed by the \gls{W2E} path is lower than that of the \gls{E2W} path.
However, as the throughput increases, the difference between the total number of cycles consumed by the \gls{W2E} and \gls{E2W} paths decreases.

In Observation \ref{obsv:core_utilization_inequalty}, we discussed that the dual-core configuration leverages both cores in the \gls{W2E} path.
By adding packets to the \qdisc{}, egress processing is offloaded to the core assigned to the Ethernet \gls{NIC} (Core 1), and the packet switching load is distributed across the two cores.
\textit{However, the processing load of the egress side processing in the \gls{W2E} path increases as more packets are added to \qdisc{} instead of being transmitted directly.}

As discussed in Section \ref{sec:reviews_tx_side}, the \qdisc{} layer acts as an intermediary between the IP layer and the network device, managing packet queues and scheduling egress transmissions.
After protocol stack processing, the \verb|__dev_xmit_skb| function (cf. Figure \ref{fig:pkt_switching_steps}) acquires a lock on the \qdisc{} and then checks if the packet can bypass the \qdisc{} under certain conditions, such as when the queue is empty.
Our further analysis (not included in this paper) revealed that for the 500 Mbps throughput, approximately 4\% of packets are added to \qdisc{}, and 96\% of them are transmitted directly.
If the \qdisc{} can be bypassed, the \verb|sch_direct_xmit| function releases the \qdisc{} lock (which was acquired by \verb|__dev_xmit_skb|) and then acquires a lock on the TX buffer.
Acquiring this lock is a process-intensive task due to two reasons.
First, the function \verb|sch_direct_xmit| running on Core 0 must compete with the \qdisc{} and the \gls{IRQ} handler of the Ethernet \gls{NIC}, both running on Core 1.
For instance, this function needs to compete with the \bcmgenet{} driver running on Core 1 when the driver has locked the TX buffer to perform operations such as reclaiming the buffer entries for the transmitted packets.
Also, on Core 1, another instance of the function \verb|sch_direct_xmit| may be running to transmit the packets out of \qdisc{}.
Note that the execution of \qdisc{} is necessary for the 4\% of packets added to the \qdisc{}, because the core assigned to the Ethernet interface is the one responsible for handling packet transmissions once those are added to the \qdisc{}.

As throughput increases, so does the number of packets that cannot bypass the \qdisc{}.
Further analysis (not included in this paper) revealed that for the maximum throughput level of the \gls{W2E} path, 76\% of packets are added to the \qdisc{}, while the rest are directly transmitted.
The increasing processing demand of the \qdisc{} is reflected in the higher number of processor cycles consumed by the \TXIRQ{} and \TXQdisc{} stages.
Specifically, comparing Figures \ref{fig:flame_graph}(b) and (c) shows that increasing the throughput from 500 to 949 Mbps results in 3x and 61x increase in the number of processor cycles consumed by these two sub-stages on the \gls{W2E} path, respectively.
These significant growths are due to two reasons. 
First, the \qdisc{} of the Ethernet interface implements the FQ-CoDel algorithm, a process-intensive algorithm that keeps track of the sojourn time of the packets belonging to each flow to avoid buffer-bloat.
Second, as discussed above, for the 24\% of packets that bypass the \qdisc{}, the system requires to compete and acquire a lock whenever access to the driver's \gls{TX} buffer is needed.

\section{Related Work and Future Opportunities}
\label{related_work}

To the best of our knowledge, there is no study on understanding and enhancing the \gls{NIC}-to-\gls{NIC} data plane of WiFi \glspl{AP}. 
The existing works related to packet switching on WiFi \glspl{AP} are primarily focused on packet scheduling, queuing, and airtime fairness \cite{hoiland2017ending,hoiland2018piece,richart2017resource,richart2020slicing,isolani2020airtime,eaps_jay,canbal2023wi,deng2022coorp}. For instance, the increased latency and jitter due to excessive queueing (bufferbloat) is discussed in \cite{hoiland2017ending}.
In another category of related works, tools have been proposed in \cite{sheth2021monfi} and \cite{sheth2021flip} to facilitate event tracing and provide additional real-time insights into the operation of the WiFi stack.
Such tools can be used in addition to the statistics provided by the kernel to better understand and enhance the kernel's data path.
It is also worth noting that this paper, as well as the studies in \cite{hoiland2017ending,hoiland2018piece,richart2017resource,richart2020slicing,isolani2020airtime,eaps_jay,canbal2023wi,deng2022coorp,sheth2021monfi,sheth2021flip}, all use SoftMAC-based WiFi \glspl{NIC}. We leave it to future work to investigate how FullMAC devices, which implement MAC functionalities in the \gls{NIC} and do not utilize the \maceighteleven module, affect system operation and performance.

There exist several works on the evaluation and enhancement of software packet switching in \textit{high-performance settings} utilizing servers with Xeon (x86-64) processors. 
The study \cite{bolla2008pc} demonstrated that binding all forwarding operations of each packet to a single core reduces cache invalidation rate and spinlock contention compared to multi-core packet switching.
A similar observation has been reported in \cite{dobrescu2009routebricks}.
Towards achieving predictable packet switching performance, \cite{dobrescu2012toward} shows that cache contention is the leading cause of performance variations. 
They also examined the potential benefits of contention-aware task scheduling and found that it provides minimal performance improvement. 
Dynamic configurations based on workload demands are emphasized in \cite{Fang:EECS-2018-136}. 
Settings such as disabling Hyper-Threading (specific to x86-64 processors) and processor core isolation are shown to be crucial for achieving high and stable performance.
The study \cite{raumer2015performance} shows that switching more than one million flows results in the considerably higher overhead of IP routing, which occurs when the routing table cannot fit entirely into the processor's cache. 
The disparity between processor utilization and the number of cycles consumed has been mentioned in \cite{emmerich2014performance}; however, they did not discuss the underlying causes of the problem.

Various works have modified the kernel path \cite{kafe2018_trans,zygos} or proposed novel algorithms and methods for the dynamic configuration of packet-switching systems \cite{meyer2014low,suksomboon2018configuring}.
To minimize delays for high-priority data flows, \cite{meyer2014low} proposes a method that involves assigning incoming packets to different RX buffers based on their priority levels and then employing various scheduling techniques to efficiently manage the resources assigned to processing incoming packets.
The problem of finding optimal configuration parameters for a software switch to achieve maximum throughput and minimum latency has been addressed in \cite{suksomboon2018configuring}.
Kafe \cite{kafe2018_trans} enhances kernel packet switching by introducing a cache-optimized SKB allocator that recycles pre-allocated buffers efficiently. This method reduces the overhead associated with frequent allocation and de-allocation of memory buffers, lowering cache misses.
The authors in \cite{zygos} proposed ZygOS, a system designed to achieve low tail latency in microsecond-scale networked tasks. ZygOS incorporates advanced interrupt handling and scheduling techniques to enhance performance. 
Given the flexibility of software switches in managing physical resources such as the number of processor cores and capacity of RX buffers, \cite{gallardo2016performance} propose a model to instantaneously estimate the minimum number of processor cores required to meet given QoS criteria.

In this paper, we leveraged several best practices, including \gls{IRQ} affinity and core isolation, to enhance switching performance. Additional optimizations are orthogonal to our work and can be employed for further performance enhancements. For instance, dynamic \gls{IRQ} affinity assignment and multi-queue \glspl{NIC} can be utilized to allocate resources based on the system's load. Furthermore, studying the impact of the number of flows and complex \gls{NFV} tasks (e.g., firewall, encryption, monitoring) on cache performance remains an area for future research.

Given the importance of communications between \gls{NIC} and \textit{user-space applications} and \glspl{VM} for cloud computing and \gls{NFV} environments, understanding and enhancing this communication has also received attention.
The study presented in \cite{bolla2008pc} demonstrated that, under conditions of high traffic, a single core dedicated to packet switching and user-space processing allocates less than 2\% of its resources to user-space tasks. 
Focused on packet switching between kernel and user-space, \cite{cai2021understanding} demonstrated that more than 50\% of the overhead is caused by packet copying operations.
The overhead of traffic switching between the user-space and two \glspl{NIC} supporting 802.11ad and 802.11ac has been studied in \cite{khan2022multipath}.
Their utilized platform has four powerful cores and four small cores, and when the \glspl{IRQ} are assigned to small cores, sub-optimal throughput is observed.
The authors of \cite{khalid2018iron} revealed that packet exchanges between \gls{NIC} and user-space applications do not accurately account for the application's processor demand.
They proposed a kernel modification of the SoftIRQ processing to address this problem.
In \cite{cena2018sdmac}, the authors provide applications with real-time access to MAC primitives, enabling direct control over packet transmissions and retransmissions at the user space level.

As WiFi \glspl{AP} continue to be deployed in a wide range of applications, it is anticipated that an increasing number of \gls{NFV} instances will run on these \glspl{AP} \cite{martinez2023wi,wu2019resource,yang2020recent,riggio2015virtual}. For example, implementing security functions like firewalls and intrusion detection directly on \glspl{AP} can provide faster threat responses, while deploying content caching applications at the edge can significantly reduce bandwidth consumption and improve user experience by delivering content more quickly. Therefore, an area of future work is to study the implications of packet switching performance for running \gls{NFV} in the kernel or user-space and to adopt various optimization methods \cite{iyer2019performance,pereira2024automatic,zhang2019comparing}.

Existing works on user-space packet switching (i.e., kernel bypass methods) demonstrate superior performance compared to Linux's layer-2 and layer-3 switching \cite{emmerich2014performance,Fang:EECS-2018-136,gallardo2016performance}. 
In \cite{emmerich2014performance}, the authors showed that for packet switching between two \glspl{NIC}, ther performance of the Linux bridge, Linux's IP forwarding, OVS-kernel and OVS-DPDK are 1.11, 1.58, 1.88 and 11.31 million packet per second, respectively.
Implementation and analysis of user-space packet switching on WiFi \glspl{AP} is left as a future work.

\section{Conclusion}
\label{sec_conclusion}

As the number of WiFi \glspl{AP} and their packet switching rate and complexity increase, it becomes increasingly important to understand and enhance the operation of \gls{AP}'s data plane.
In this paper, we presented a thorough analysis of packet switching on WiFi \glspl{AP} and targeted the following goals.
The first goal is to rely on the statistics provided by the Linux kernel to understand the operation of the kernel's data paths for both \gls{W2E} and \gls{E2W} packet switching.
In this regard, we showed that several factors can affect the meaning and reliability of these statistics.
For instance, the kernel may not accurately account for the SoftIRQs executed within the \gls{IRQ} context, the TX \glspl{IRQ} may contribute to the number of RX SoftIRQs, and device drivers may override \gls{NAPI} operation.
The second goal is to understand the differences in the two data paths, namely, \gls{E2W} and \gls{W2E}, regarding the components along each path and their overhead under various throughput loads.
We showed that packet switching through these paths exhibits significant differences that affect core utilization and energy efficiency.

We hope that these findings pave the way towards understanding and enhancing the operation of \glspl{AP}. An area of future work includes dynamic system configuration, such as \gls{IRQ} affinity assignment, based on the system's real-time load.
Further study and improvement of \glspl{AP} operation using techniques such as cache-aware \gls{NFV}, user-space packet switching, multi-queue drivers (with multiple RX and TX buffers), and \gls{RPS} are future research areas.

It is worth noting that in this work we used the Linux operating system on a certain type of hardware.
While our primary aim is to identify overarching trends and establish general principles, it is important to acknowledge that different operating systems (e.g., different Linux kernel versions) and device drivers (e.g., SoftMAC and FullMAC) may demonstrate varying operational and performance characteristics.

\section*{Acknowledgment}
This work was supported by NSF grant \#2138633 and an internal grant from Santa Clara University's Center for Sustainability.
The authors would like to thank Netgear for donating some of the materials used to conduct this research.

\ifCLASSOPTIONcaptionsoff
  \newpage
\fi

\bibliographystyle{IEEEtran}

\bibliography{references}

\begin{thebibliography}{10}
\providecommand{\url}[1]{#1}
\csname url@samestyle\endcsname
\providecommand{\newblock}{\relax}
\providecommand{\bibinfo}[2]{#2}
\providecommand{\BIBentrySTDinterwordspacing}{\spaceskip=0pt\relax}
\providecommand{\BIBentryALTinterwordstretchfactor}{4}
\providecommand{\BIBentryALTinterwordspacing}{\spaceskip=\fontdimen2\font plus
\BIBentryALTinterwordstretchfactor\fontdimen3\font minus
  \fontdimen4\font\relax}
\providecommand{\BIBforeignlanguage}[2]{{%
\expandafter\ifx\csname l@#1\endcsname\relax
\typeout{** WARNING: IEEEtran.bst: No hyphenation pattern has been}%
\typeout{** loaded for the language `#1'. Using the pattern for}%
\typeout{** the default language instead.}%
\else
\language=\csname l@#1\endcsname
\fi
#2}}
\providecommand{\BIBdecl}{\relax}
\BIBdecl

\bibitem{reshef2022future}
E.~Reshef and C.~Cordeiro, ``\relax{Future directions for Wi-Fi 8 and
  beyond},'' \emph{IEEE Communications Magazine}, vol.~60, no.~10, pp. 50--55,
  2022.

\bibitem{wifi_ap_market_2024}
\BIBentryALTinterwordspacing
F.~M. Intelligence, ``\relax{Gigabit Wi-Fi Access Point Market Outlook for 2024
  to 2034}.''\hskip 1em plus 0.5em minus 0.4em\relax FMI, 2022. [Online].
  Available:
  \url{https://www.futuremarketinsights.com/reports/gigabit-wi-fi-access-point-market}
\BIBentrySTDinterwordspacing

\bibitem{hoiland2017ending}
T.~H{\o}iland-J{\o}rgensen, M.~Kazior, D.~T{\"a}ht, P.~Hurtig, and
  A.~Brunstrom, ``\relax{Ending the Anomaly: Achieving Low Latency and Airtime
  Fairness in WiFi},'' in \emph{USENIX Annual Technical Conference}, 2017, pp.
  139--151.

\bibitem{sheth2021monfi}
J.~Sheth and B.~Dezfouli, ``Monfi: A tool for high-rate, efficient, and
  programmable monitoring of wifi devices,'' in \emph{IEEE Wireless
  Communications and Networking Conference (WCNC)}.\hskip 1em plus 0.5em minus
  0.4em\relax IEEE, 2021, pp. 1--7.

\bibitem{eaps_jay}
J.~Sheth, C.~Miremadi, A.~Dezfouli, and B.~Dezfouli, ``Eaps: Edge-assisted
  predictive sleep scheduling for 802.11 iot stations,'' \emph{IEEE Systems
  Journal}, vol.~16, no.~1, pp. 591--602, 2022.

\bibitem{emmerich2014performance}
P.~Emmerich, D.~Raumer, F.~Wohlfart, and G.~Carle, ``Performance
  characteristics of virtual switching,'' in \emph{IEEE 3rd International
  Conference on Cloud Networking (CloudNet)}, 2014, pp. 120--125.

\bibitem{powell2020fog}
C.~Powell, C.~Desiniotis, and B.~Dezfouli, ``The fog development kit: A
  platform for the development and management of fog systems,'' \emph{IEEE
  Internet of Things Journal}, vol.~7, no.~4, pp. 3198--3213, 2020.

\bibitem{cai2021understanding}
Q.~Cai, S.~Chaudhary, M.~Vuppalapati, J.~Hwang, and R.~Agarwal, ``Understanding
  host network stack overheads,'' in \emph{Proceedings of the ACM SIGCOMM},
  2021, pp. 65--77.

\bibitem{okelmann2021adaptive}
P.~Okelmann, L.~Linguaglossa, F.~Geyer, P.~Emmerich, and G.~Carle, ``Adaptive
  batching for fast packet processing in software routers using machine
  learning,'' in \emph{IEEE 7th International Conference on Network
  Softwarization (NetSoft)}.\hskip 1em plus 0.5em minus 0.4em\relax IEEE, 2021,
  pp. 206--210.

\bibitem{zhang2019comparing}
T.~Zhang, L.~Linguaglossa, M.~Gallo, P.~Giaccone, L.~Iannone, and J.~Roberts,
  ``\relax{Comparing the performance of state-of-the-art software switches for
  NFV},'' in \emph{CoNEXT}, 2019, pp. 68--81.

\bibitem{gallardo2016performance}
G.~A. Gallardo, B.~Baynat, and T.~Begin, ``Performance modeling of virtual
  switching systems,'' in \emph{IEEE 24th International Symposium on Modeling,
  Analysis and Simulation of Computer and Telecommunication Systems
  (MASCOTS)}.\hskip 1em plus 0.5em minus 0.4em\relax IEEE, 2016, pp. 125--134.

\bibitem{chen2021predictable}
J.~Chen and B.~Dezfouli, ``Predictable bandwidth slicing with open vswitch,''
  in \emph{IEEE Global Communications Conference (GLOBECOM)}, 2021, pp. 1--6.

\bibitem{Qualcomm_IPQ8074}
\BIBentryALTinterwordspacing
``\relax{IPQ8074},'' n.d., accessed: 2024-04-5. [Online]. Available:
  \url{https://www.qualcomm.com/products/internet-of-things/networking/wi-fi-networks/ipq8074}
\BIBentrySTDinterwordspacing

\bibitem{Qualcomm_QCS5430}
\BIBentryALTinterwordspacing
Qualcomm, ``\relax{QCS5430},'' n.d., accessed: 2024-04-5. [Online]. Available:
  \url{https://www.qualcomm.com/products/internet-of-things/industrial/industrial-automation/qcs5430}
\BIBentrySTDinterwordspacing

\bibitem{BCM54210}
\BIBentryALTinterwordspacing
Broadcom, ``\relax{BCM5420: Single Port RGMII SGMII Gigabit Ethernet
  Transceiver},'' n.d., accessed: 2024-04-10. [Online]. Available:
  \url{https://www.broadcom.com/products/ethernet-connectivity/phy-and-poe/copper/gigabit/bcm54210}
\BIBentrySTDinterwordspacing

\bibitem{dezfouli2018empiot}
B.~Dezfouli, I.~Amirtharaj, and C.-C. Li, ``\relax{EMPIOT: An energy
  measurement platform for wireless IoT devices},'' \emph{Journal of Network
  and Computer Applications}, vol. 121, pp. 135--148, 2018.

\bibitem{ftrace_utility}
\BIBentryALTinterwordspacing
\relax{The Linux Kernel}, ``\relax{ftrace - Function Tracer},'' n.d., accessed:
  2024-02-10. [Online]. Available:
  \url{https://www.kernel.org/doc/html/v6.0/trace/ftrace.html}
\BIBentrySTDinterwordspacing

\bibitem{wifi-ap-github}
\BIBentryALTinterwordspacing
\relax{SIOTLAB}, ``\relax{Understanding and Enhancing Linux Kernel-based Packet
  Switching on WiFi Access Points},'' 2024. [Online]. Available:
  \url{https://github.com/SIOTLAB/wifi-packet-switching-analysis}
\BIBentrySTDinterwordspacing

\bibitem{rothberg2015interrupt}
V.~Rothberg, ``\relax{Interrupt handling in Linux},'' University of Erlangen,
  Germany, Tech. Rep. CS-2015-07, November 2015.

\bibitem{iwlwifiWiki}
{Linux Wireless}, ``iwlwifi,'' \url{https://wireless.wiki.kernel.org/en/users
  /drivers/ iwlwifi}, n.d., accessed: 2024-03-12.

\bibitem{realtek_rtw89_threadedIRQ}
\relax{Realtek Corporation}, ``A repo for the newest realtek rtw89 codes,''
  \url{https://github.com/lwfinger/rtw89}, n.d., accessed: 2024-04-3.

\bibitem{hoiland2018piece}
T.~H{\o}iland-J{\o}rgensen, D.~T{\"a}ht, and J.~Morton, ``Piece of cake: a
  comprehensive queue management solution for home gateways,'' in \emph{IEEE
  International Symposium on Local and Metropolitan Area Networks
  (LANMAN)}.\hskip 1em plus 0.5em minus 0.4em\relax IEEE, 2018, pp. 37--42.

\bibitem{richart2017resource}
M.~Richart, J.~Baliosian, J.~Serrati, J.-L. Gorricho, R.~Ag{\"u}ero, and
  N.~Agoulmine, ``\relax{Resource allocation for network slicing in WiFi access
  points},'' in \emph{13th International conference on network and service
  management (CNSM)}.\hskip 1em plus 0.5em minus 0.4em\relax IEEE, 2017, pp.
  1--4.

\bibitem{richart2020slicing}
M.~Richart, J.~Baliosian, J.~Serrat, J.-L. Gorricho, and R.~Ag{\"u}ero,
  ``\relax{Slicing with guaranteed quality of service in WiFi networks},''
  \emph{IEEE Transactions on Network and Service Management}, vol.~17, no.~3,
  pp. 1822--1837, 2020.

\bibitem{isolani2020airtime}
P.~H. Isolani, N.~Cardona, C.~Donato, G.~A. P{\'e}rez, J.~M. Marquez-Barja,
  L.~Z. Granville, and S.~Latr{\'e}, ``\relax{Airtime-based resource allocation
  modeling for network slicing in IEEE 802.11 RANs},'' \emph{IEEE
  Communications Letters}, vol.~24, no.~5, pp. 1077--1080, 2020.

\bibitem{canbal2023wi}
F.~Canbal, Y.~B. Ozgun, M.~S. Kuran, G.~Venkatesan, and N.~Canpolat, ``Wi-fi
  qos management program: Bridging the qos gap of multimedia traffic in wi-fi
  networks,'' \emph{IEEE Communications Magazine}, 2023.

\bibitem{deng2022coorp}
S.~Deng, X.~Guan, Z.~Sun, S.~Zhao, T.~Shen, X.~Chen, T.~Duan, Y.~Wang, J.~Pan,
  Y.~Wu \emph{et~al.}, ``Coorp: Satisfying low-latency and high-throughput
  requirements of wireless network for coordinated robotic learning,''
  \emph{IEEE Internet of Things Journal}, vol.~10, no.~3, pp. 1946--1960, 2022.

\bibitem{sheth2021flip}
J.~Sheth, V.~Ramanna, and B.~Dezfouli, ``\relax{Flip: A framework for
  leveraging ebpf to augment WiFi access points and investigate network
  performance},'' in \emph{Proceedings of the 19th ACM International Symposium
  on Mobility Management and Wireless Access}, 2021, pp. 117--125.

\bibitem{bolla2008pc}
R.~Bolla and R.~Bruschi, ``\relax{PC-based software routers: High performance
  and application service support},'' in \emph{Proceedings of the ACM workshop
  on Programmable routers for extensible services of tomorrow}, 2008, pp.
  27--32.

\bibitem{dobrescu2009routebricks}
M.~Dobrescu, N.~Egi, K.~Argyraki, B.-G. Chun, K.~Fall, G.~Iannaccone, A.~Knies,
  M.~Manesh, and S.~Ratnasamy, ``Routebricks: Exploiting parallelism to scale
  software routers,'' in \emph{ACM SIGOPS 22nd symposium on Operating systems
  principles}, 2009, pp. 15--28.

\bibitem{dobrescu2012toward}
M.~Dobrescu, K.~Argyraki, and S.~Ratnasamy, ``Toward predictable performance in
  software packet-processing platforms,'' in \emph{USENIX Symposium on
  Networked Systems Design and Implementation (NSDI)}, 2012, pp. 141--154.

\bibitem{Fang:EECS-2018-136}
V.~Fang, T.~Lévai, S.~Han, S.~Ratnasamy, B.~Raghavan, and J.~Sherry,
  ``Evaluating software switches: Hard or hopeless?'' EECS Department,
  University of California, Berkeley, Tech. Rep. UCB/EECS-2018-136, Oct 2018.

\bibitem{raumer2015performance}
D.~Raumer, F.~Wohlfart, D.~Scholz, P.~Emmerich, and G.~Carle, ``Performance
  exploration of software-based packet processing systems,'' \emph{Leistungs-,
  Zuverl{\"a}ssigkeits-und Verl{\"a}sslichkeitsbewertung von
  Kommunikationsnetzen und verteilten Systemen}, vol.~8, 2015.

\bibitem{kafe2018_trans}
C.-H. Hong, K.~Lee, J.~Hwang, H.~Park, and C.~Yoo, ``Kafe: Can os kernels
  forward packets fast enough for software routers?'' \emph{IEEE/ACM
  Transactions on Networking}, vol.~26, no.~6, pp. 2734--2747, 2018.

\bibitem{zygos}
G.~Prekas, M.~Kogias, and E.~Bugnion, ``Zygos: Achieving low tail latency for
  microsecond-scale networked tasks,'' in \emph{Proceedings of the 26th
  Symposium on Operating Systems Principles (SOSP)}, 2017, p. 325–341.

\bibitem{meyer2014low}
T.~Meyer, D.~Raumer, F.~Wohlfart, B.~E. Wolfinger, and G.~Carle, ``Low latency
  packet processing in software routers,'' in \emph{International Symposium on
  Performance Evaluation of Computer and Telecommunication Systems
  (SPECTS)}.\hskip 1em plus 0.5em minus 0.4em\relax IEEE, 2014, pp. 556--563.

\bibitem{suksomboon2018configuring}
K.~Suksomboon, N.~Matsumoto, S.~Okamoto, M.~Hayashi, and Y.~Ji, ``Configuring a
  software router by the erlang-$k$-based packet latency prediction,''
  \emph{IEEE Journal on Selected Areas in Communications}, vol.~36, no.~3, pp.
  422--437, 2018.

\bibitem{khan2022multipath}
I.~Khan, M.~Ghoshal, S.~Aggarwal, D.~Koutsonikolas, and J.~Widmer, ``Multipath
  tcp in smartphones equipped with millimeter wave radios,'' in
  \emph{Proceedings of the 15th ACM Workshop on Wireless Network Testbeds,
  Experimental evaluation \& Characterization}, 2022, pp. 54--60.

\bibitem{khalid2018iron}
J.~Khalid, E.~Rozner, W.~Felter, C.~Xu, K.~Rajamani, A.~Ferreira, and
  A.~Akella, ``Iron: Isolating network-based cpu in container environments,''
  in \emph{15th USENIX Symposium on Networked Systems Design and Implementation
  (NSDI)}, 2018, pp. 313--328.

\bibitem{cena2018sdmac}
G.~Cena, S.~Scanzio, and A.~Valenzano, ``\relax{SDMAC: a software-defined MAC
  for Wi-Fi to ease implementation of soft real-time applications},''
  \emph{IEEE Transactions on Industrial Informatics}, vol.~15, no.~6, pp.
  3143--3154, 2018.

\bibitem{martinez2023wi}
V.~M. Mart{\'\i}nez, M.~R. Ribeiro, and V.~F. Mota, ``\relax{Wi-Fi faces the
  new wireless ecosystem: a critical review},'' \emph{Annals of
  Telecommunications}, pp. 1--17, 2023.

\bibitem{wu2019resource}
B.~Wu, J.~Zeng, L.~Ge, S.~Shao, Y.~Tang, and X.~Su, ``\relax{Resource
  allocation optimization in the NFV-enabled MEC network based on game
  theory},'' in \emph{IEEE International Conference on Communications
  (ICC)}.\hskip 1em plus 0.5em minus 0.4em\relax IEEE, 2019, pp. 1--7.

\bibitem{yang2020recent}
S.~Yang, F.~Li, S.~Trajanovski, R.~Yahyapour, and X.~Fu, ``Recent advances of
  resource allocation in network function virtualization,'' \emph{IEEE
  Transactions on Parallel and Distributed Systems}, vol.~32, no.~2, pp.
  295--314, 2020.

\bibitem{riggio2015virtual}
R.~Riggio, T.~Rasheed, and R.~Narayanan, ``Virtual network functions
  orchestration in enterprise wlans,'' in \emph{IFIP/IEEE International
  Symposium on Integrated Network Management (IM)}.\hskip 1em plus 0.5em minus
  0.4em\relax IEEE, 2015, pp. 1220--1225.

\bibitem{iyer2019performance}
R.~Iyer, L.~Pedrosa, A.~Zaostrovnykh, S.~Pirelli, K.~Argyraki, and G.~Candea,
  ``Performance contracts for software network functions,'' in \emph{USENIX
  Symposium on Networked Systems Design and Implementation (NSDI)}, 2019, pp.
  517--530.

\bibitem{pereira2024automatic}
F.~Pereira, F.~M. Ramos, and L.~Pedrosa, ``Automatic parallelization of
  software network functions,'' in \emph{USENIX Symposium on Networked Systems
  Design and Implementation (NSDI)}, 2024, pp. 1531--1550.

\end{thebibliography}

\balance

\end{document}